\documentclass[10pt, journal]{IEEEtran}
\usepackage{makecell}
\usepackage{url}
\usepackage[utf8]{inputenc}
\usepackage[dvipsnames]{xcolor}
\usepackage{amsmath, amssymb}
\usepackage{pifont}
\newcommand{\cmark}{\ding{51}}%
\newcommand{\xmark}{\ding{55}}%
\usepackage{dsfont}

\usepackage{microtype}
\usepackage[acronyms,nonumberlist,nopostdot,nomain,nogroupskip]{glossaries}
\usepackage{tablefootnote}
\usepackage{booktabs}
\usepackage{tabularx}
\usepackage{epsfig}
\usepackage[outdir=img/]{epstopdf}
\usepackage{tikz}
\usepackage{soul}
\usepackage{pgfplots}
\pgfplotsset{compat=newest} 
\pgfplotsset{plot coordinates/math parser=false} 
\newlength\fheight
\newlength\fwidth
\usetikzlibrary{plotmarks,shapes,patterns,decorations.pathreplacing,backgrounds,calc,arrows,arrows.meta,spy,matrix,shadows,trees,positioning, datavisualization,datavisualization.formats.functions}
\usepgfplotslibrary{patchplots,groupplots, fillbetween, colorbrewer}
\usepackage{tikzscale}
\usepackage{tikz-qtree}
\usepackage{hyperref}
\usepackage{siunitx}
\usepackage{bm}
\usepackage{pgfplotstable}
\usepackage[multiuser]{fixme}
\fxsetup{status=draft, envlayout=color, theme=color}
\definecolor{fxtarget}{rgb}{1.,0.0000,0.0000}
\FXRegisterAuthor{mb}{amb}{MB}
\usepackage{multirow}

\renewcommand{\figurename}{Fig.}
\usepackage[font=scriptsize]{subcaption}
\usepackage[font=footnotesize]{caption}

\usepackage{mathtools}

\usepackage{dblfloatfix}    
\usepackage{colortbl}
\usepackage{flushend}

\newacronym{quic}{QUIC}{Quick UDP Internet Connections}
\newacronym{3gpp}{3GPP}{3rd Generation Partnership Project}
\newacronym{adc}{ADC}{Analog to Digital Converter}
\newacronym{5g}{5G}{5th Generation}
\newacronym{aimd}{AIMD}{Additive Increase Multiplicative Decrease}
\newacronym{am}{AM}{Acknowledged Mode}
\newacronym{amc}{AMC}{Adaptive Modulation and Coding}
\newacronym{aqm}{AQM}{Active Queue Management}
\newacronym{awgn}{AGWN}{Additive White Gaussian Noise}
\newacronym{balia}{BALIA}{Balanced Link Adaptation}
\newacronym{bdp}{BDP}{Bandwidth-Delay Product}
\newacronym{bf}{BF}{Beamforming}
\newacronym{cc}{CC}{Congestion Control}
\newacronym{pdf}{PDF}{Probability Density Function}
\newacronym{cdf}{CDF}{Cumulative Distribution Function}
\newacronym{icdf}{Inverse CDF}{Inverse Cumulative Distribution Function}
\newacronym{c-v2x}{C-V2X}{Cellular Vehicle-To-Everything}
\newacronym{ci}{CI}{Close-in free space reference}
\newacronym{cn}{CN}{Core Network}
\newacronym{cqi}{CQI}{Channel Quality Information}
\newacronym{cp}{CP}{Control Plane}
\newacronym{csirs}{CSI-RS}{Channel State Information - Reference Signal}
\newacronym{d2d}{D2D}{Device-to-Device}
\newacronym{dc}{DC}{Dual Connectivity}
\newacronym{mle}{MLE}{Maximum Likelihood Estimation}
\newacronym{dce}{DCE}{Direct Code Execution}
\newacronym{dci}{DCI}{Downlink Control Information}
\newacronym{dmr}{DMR}{Deadline Miss Ratio}
\newacronym{dmrs}{DMRS}{DeModulation Reference Signal}
\newacronym{dsrc}{DSRC}{Dedicated Short-Range Communication}
\newacronym{e2e}{E2E}{End-to-End}
\newacronym{ecn}{ECN}{Explicit Congestion Notification}
\newacronym{edf}{EDF}{Earliest Deadline First}
\newacronym{enb}{eNB}{evolved Node Base}
\newacronym{epc}{EPC}{Evolved Packet Core}
\newacronym{es}{ES}{Edge Server}
\newacronym{fdma}{FDMA}{Frequency Division Multiple Access}
\newacronym{fdd}{FDD}{Frequency Division Duplexing}
\newacronym[firstplural=Radio Access Technologies (RATs)]{rat}{RAT}{Radio Access Technology}
\newacronym{fs}{FS}{Fast Switching}
\newacronym{ftp}{FTP}{File Transfer Protocol}
\newacronym{gnb}{gNB}{Next Generation Node Base}
\newacronym{harq}{HARQ}{Hybrid Automatic Repeat reQuest}
\newacronym{hetnet}{HetNet}{Heterogeneous Network}
\newacronym{hh}{HH}{Hard Handover}
\newacronym{hol}{HOL}{Head-of-Line}
\newacronym{ia}{IA}{Initial Access}
\newacronym{imt}{IMT}{International Mobile Telecommunication}
\newacronym{iot}{IoT}{Internet of Things}
\newacronym{lidar}{LiDAR}{Light Detection and Ranging}
\newacronym{los}{LoS}{Line of Sight}
\newacronym{lstm}{LSTM}{long short-term memory}
\newacronym{lte}{LTE}{Long Term Evolution}
\newacronym{m2m}{M2M}{Machine-to-Machine}
\newacronym{mac}{MAC}{Medium Access Control}
\newacronym{mc}{MC}{Multi-Connectivity}
\newacronym{mcs}{MCS}{Modulation and Coding Scheme}
\newacronym{mec}{MEC}{Mobile Edge Cloud}
\newacronym{mi}{MI}{Mutual Information}
\newacronym{mimo}{MIMO}{Multiple Input, Multiple Output}
\newacronym{mmwave}{mmWave}{millimeter wave}
\newacronym{ml}{ML}{Machine Learning}
\newacronym{mpc}{MPC}{Model Predictive Control}
\newacronym{mr}{MR}{Maximum Rate}
\newacronym{mss}{MSS}{Maximum Segment Size}
\newacronym{mtd}{MTD}{Machine-Type Device}
\newacronym{mtu}{MTU}{Maximum Transmission Unit}
\newacronym{nn}{NN}{Neural Network}
\newacronym{nsf}{NSF}{National Science Foundation}
\newacronym{nfv}{NFV}{Network Function Virtualization}
\newacronym{nlos}{nLoS}{non-Line-of-Sight}
\newacronym{nr}{NR}{New Radio}
\newacronym{ofdm}{OFDM}{Orthogonal Frequency Division Multiplexing}
\newacronym{pc}{PC}{Point Cloud}
\newacronym{pdcch}{PDCCH}{Physical Downlonk Control Channel}
\newacronym{pdcp}{PDCP}{Packet Data Convergence Protocol}
\newacronym{pdsch}{PDSCH}{Physical Downlink Shared Channel}
\newacronym{pdu}{PDU}{Packet Data Unit}
\newacronym{pf}{PF}{Proportional Fair}
\newacronym{pgw}{PGW}{Packet Gateway}
\newacronym{phy}{PHY}{Physical}
\newacronym{pbch}{PBCH}{Physical Broadcast Channel}
\newacronym[plural=\gls{mme}s,firstplural=Mobility Management Entities (MMEs)]{mme}{MME}{Mobility Management Entity}
\newacronym{prb}{PRB}{Physical Resource Block}
\newacronym{pss}{PSS}{Primary Synchronization Signal}
\newacronym{pucch}{PUCCH}{Physical Uplink Control Channel}
\newacronym{pusch}{PUSCH}{Physical Uplink Shared Channel}
\newacronym{qos}{QOS}{Quality of Service}
\newacronym{rach}{RACH}{Random Access Channel}
\newacronym{ran}{RAN}{Radio Access Network}
\newacronym{red}{RED}{Random Early Detection}
\newacronym{rf}{RF}{Radio Frequency}
\newacronym{rlc}{RLC}{Radio Link Control}
\newacronym{rlf}{RLF}{Radio Link Failure}
\newacronym{rrc}{RRC}{Radio Resource Control}
\newacronym{rrm}{RRM}{Radio Resource Management}
\newacronym{rr}{RR}{Round Robin}
\newacronym{rs}{RS}{Remote Server}
\newacronym{rsrp}{RSRP}{Reference Signal Received Power}
\newacronym{rss}{RSS}{Received Signal Strength}
\newacronym{rtt}{RTT}{Round Trip Time}
\newacronym{rw}{RW}{Receive Window}
\newacronym{rx}{RX}{Receiver}
\newacronym{sa}{SA}{standalone}
\newacronym{sack}{SACK}{Selective Acknowledgment}
\newacronym{sap}{SAP}{Service Access Point}
\newacronym{sch}{SCH}{Secondary Cell Handover}
\newacronym{scoot}{SCOOT}{Split Cycle Offset Optimization Technique}
\newacronym{sdma}{SDMA}{Spatial Division Multiple Access}
\newacronym{sinr}{SINR}{Signal to Interference plus Noise Ratio}
\newacronym{sm}{SM}{Saturation Mode}
\newacronym{snr}{SNR}{Signal to Noise Ratio}
\newacronym{psnr}{PSNR}{Peak Signal to Noise Ratio}
\newacronym{son}{SON}{Self-Organizing Network}
\newacronym{ss}{SS}{Synchronization Signal}
\newacronym{srs}{SRS}{Sounding Reference Signal}
\newacronym{sss}{SSS}{Secondary Synchronization Signal}
\newacronym{tb}{TB}{Transport Block}
\newacronym{tcp}{TCP}{Transmission Control Protocol}
\newacronym{tdd}{TDD}{Time Division Duplexing}
\newacronym{tdma}{TDMA}{Time Division Multiple Access}
\newacronym{tfl}{TfL}{Transport for London}
\newacronym{thz}{THz}{Terahertz}
\newacronym{tm}{TM}{Transparent Mode}
\newacronym{trp}{TRP}{Transmitter Receiver Pair}
\newacronym{tti}{TTI}{Transmission Time Interval}
\newacronym{ttt}{TTT}{Time-to-Trigger}
\newacronym{tx}{TX}{Transmitter}
\newacronym{ue}{UE}{User Equipment}
\newacronym{ul}{UL}{Uplink}
\newacronym{uml}{UML}{Unified Modeling Language}
\newacronym{um}{UM}{Unacknowledged Mode}
\newacronym{utc}{UTC}{Urban Traffic Control}
\newacronym{v2i}{V2I}{Vehicle-to-Infrastructure}
\newacronym{v2v}{V2V}{Vehicle-to-Vehicle}
\newacronym{vm}{VM}{Virtual Machine}
\newacronym{rsrq}{RSRQ}{Reference Signal Received Quality}
\newacronym{rssi}{RSSI}{Received Signal Strength Indicator}
\newacronym{crs}{CRS}{Cell Reference Signal}
\newacronym{comp}{CoMP}{Coordinated Multi-Point}
\newacronym{cran}{C-RAN}{Cloud \acrlong{ran}}
\newacronym{ca}{CA}{Carrier Aggregation}
\newacronym{cco}{CC}{Carrier Component}
\newacronym{nsa}{NSA}{Non Stand Alone}
\newacronym{embb}{eMBB}{Enhanced Mobility Broadband}
\newacronym{bsr}{BSR}{Buffer Status Report}
\newacronym{srb}{SRB}{Service Radio Bearer}
\newacronym{scm}{SCM}{Spatial Channel Model}
\newacronym{sctp}{SCTP}{Stream Control Transmission Protocol}
\newacronym{mptcp}{MPTCP}{Multi-path TCP}
\newacronym{ietf}{IETF}{Internet Engineering Task Force}
\newacronym{os}{OS}{Operating System}
\newacronym{tls}{TLS}{Transport Layer Security}
\newacronym{rfc}{RFC}{Request for Comments}
\newacronym{http}{HTTP}{HyperText Transfer Protocol}
\newacronym{nat}{NAT}{Network Address Translation}
\newacronym{api}{API}{Application Programming Interface}
\newacronym{rto}{RTO}{Retransmission Timeout}
\newacronym{psc}{PSC}{Public Safety Communication}
\newacronym{rpgm}{RPGM}{Reference Point Group Mobility}
\newacronym{ic}{IC}{Incident Command}
\newacronym{rsu}{RSU}{Road Side Unit}
\newacronym{uav}{UAV}{Unmanned Aerial Vehicle}
\newacronym{usa}{U.S.}{United States}
\newacronym{vr}{VR}{Virtual Reality}
\newacronym{iab}{IAB}{Integrated Access and Backhaul}
\newacronym{wlan}{WLAN}{Wireless Local Area Network}
\newacronym{cots}{COTS}{Commercial Off-the-Shelf}
\newacronym{fpga}{FPGA}{Field Programmable Gate Array}
\newacronym{rcn}{RCN}{Research Coordination Network}
\newacronym{abg}{ABG}{Alpha-Beta-Gamma}
\newacronym{fi}{FI}{Floating Intercept}
\newacronym{uas}{UAS}{Unmanned Aerial System}
\newacronym{gps}{GPS}{Global Positioning System}
\newacronym{a2g}{A2G}{air-to-ground}
\newacronym{a2a}{A2A}{air-to-air}
\newacronym{uma}{UMa}{Urban Macro}
\newacronym{umi}{UMi}{Urban Micro}
\newacronym{upa}{UPA}{Uniform Planar Array}
\newacronym{rma}{RMa}{Rural Macro}
\newacronym{inoo}{InOo}{Indoor Open Office}
\newacronym{ple}{PLE}{path loss exponent}
\newacronym{aoa}{AoA}{Angle of Arrival}
\newacronym{aod}{AoD}{Angle of Departure}
\newacronym{toa}{ToA}{Time of Arrival}
\newacronym{cir}{CIR}{Channel Impulse Response}
\newacronym{rt}{RT}{Ray-tracing}
\newacronym{tc}{TC}{Time Cluster}
\newacronym{svd}{SVD}{Singular Value Decomposition}
\newacronym{6g}{6G}{sixth generation}
\newacronym{ns3}{ns-3}{Network Simulator 3}
\newacronym{fsc}{FS}{Fully Stochastic}
\newacronym{hbc}{HB}{Hybrid}
\newacronym{hpbw}{HPBW}{Half Power Beamwidth}
\newacronym{hsc}{HSC}{Hybrid Semantic Compression}
\newacronym{prr}{PRR}{Packet Receipt Rate}
\newacronym{v2x}{V2X}{Vehicle-To-Everything}
\newacronym{dl}{DL}{Deep Learning}
\newacronym{cl}{CL}{Compression Level}
\newacronym{sl}{SL}{Semantic Level}
\newacronym{qp}{QP}{Quantization Parameter}
\newacronym{p2pPSNR}{p2p-PSNR}{Point-to-Plane PSNR}
\newacronym{p2p}{P2P}{Peer-to-Peer}
\newacronym{xr}{XR}{eXtended Reality}
\newacronym{hmd}{HMD}{Head Mounted Device}
\newacronym{ks}{KS}{Kolmogorov-Smirnoff}
\newacronym{nrmse}{NRMSE}{Normalized Root Mean Squared Error}

\newif\ifexttikz
\exttikzfalse

\ifexttikz
	\usetikzlibrary{external}
\fi

\usepackage{multirow}
\usepackage{enumerate}
\usetikzlibrary {patterns.meta}
\renewcommand{\figurename}{Fig.}

\usepackage[capitalize]{cleveref}
\crefname{section}{Sec.}{Secs.}

\usepackage{mathtools}

\def\BibTeX{{\rm B\kern-.05em{\sc i\kern-.025em b}\kern-.08em
    T\kern-.1667em\lower.7ex\hbox{E}\kern-.125emX}}

\makeatletter
\newcommand*\titleheader[1]{\gdef\@titleheader{#1}}
\AtBeginDocument{%
  \let\st@red@title\@title
  \def\@title{%
    \bgroup\normalfont\small\centering\@titleheader\par\egroup
    \vskip1.5em\st@red@title}
}
\makeatother

\titleheader{
This work has been submitted to the IEEE for possible publication.  Copyright may be transferred without notice, after which this version may no longer be accessible.
}
\title{Statistical Analysis and End-to-End Performance Evaluation of Traffic Models for Automotive Data}
\author{Marcello Bullo,~\IEEEmembership{Student~Member,~IEEE}, Amir Ashtari Gargari, Paolo Testolina,~\IEEEmembership{Member,~IEEE}, \\
Michele~Zorzi,~\IEEEmembership{Fellow,~IEEE},
Marco Giordani,~\IEEEmembership{Member,~IEEE}

\thanks{
Marcello Bullo was with the Department of Information Engineering (DEI) of the University of Padova, Italy. He is now with the Department of Electrical and Electronic Engineering, Imperial College London, UK. Email: m.bullo21@imperial.ac.uk.

Amir Ashtari Gargari was with the Department of Information Engineering (DEI) of the University of Padova, Italy. He is now with the Centre Tecnologic de Telecomunicacions de Catalunya (CTTC), Barcelona, Spain. Email: amir.ashtari@cttc.cat.

Paolo Testolina was with the Department of Information Engineering (DEI) of the University of Padova, Italy. He is now with Northeastern University, Boston, MA, USA. Email: p.testolina@northeastern.edu.

Michele Zorzi and Marco Giordani are with the Department of Information Engineering (DEI) of the University of Padova, Italy. Email: \{giordani,zorzi\}@dei.unipd.it.

This work received funding from UKRI (Grant No. EP/X030806/1), and it was partially supported by the European Union under the Italian National Recovery and Resilience Plan (NRRP) of NextGenerationEU, partnership on “Telecommunications of the Future” (PE0000001 - program “RESTART”).
The work of P.\ Testolina was partially supported by Fondazione CaRiPaRo under grants ``Dottorati di Ricerca'' 2019.

}}

\begin{document}

\maketitle

\begin{abstract}

    Autonomous driving is a major paradigm shift in transportation, with the potential to enhance safety, optimize traffic congestion, and reduce fuel consumption. Although autonomous vehicles rely on advanced sensors and on-board computing systems to navigate without human control, full awareness of the driving environment also requires a cooperative effort via \gls{v2x} communication. Specifically, vehicles send and receive sensor perceptions to/from other vehicles to extend perception beyond their own sensing range. However, transmitting large volumes of data can be challenging for current \gls{v2x} communication technologies, so data compression represents a crucial solution to reduce the message size and link~congestion.

    In this paper, we present a statistical characterization of automotive data, focusing on \gls{lidar} sensors. 
    Notably, we provide models for the size of both raw and compressed point clouds. 
    The use of statistical traffic models offers several advantages compared to using real data, such as faster simulations, reduced storage requirements, and greater flexibility in the application design. 
    Furthermore, statistical models can be used for understanding traffic patterns and analyzing statistics, which is crucial to design and optimize wireless networks.
    We validate our statistical models via a \gls{ks} test implementing a Bootstrap Resampling scheme. Moreover, we show via ns-3 simulations that using statistical models yields comparable results in terms of latency and throughput compared to real data, which also demonstrates the accuracy of the models.

\end{abstract}

\glsresetall


\section{Introduction}
\IEEEPARstart{A}{utonomous} driving is expected to play a critical role in the development of future \gls{6g} wireless networks~\cite{giordani20196g}, redefining the way we perceive, interact with, and utilize vehicles on the roads. 
This paradigm will reduce accidents, improve the traffic flow, decrease the fuel consumption, and provide newfound mobility options for individuals with disabilities and older people~\cite{fagnant2015preparing}.

Unlike conventional vehicles where a human driver takes control, autonomous vehicles will be equipped with sensors 
and powerful computing units to perceive the environment, make driving decisions, and navigate autonomously~\cite{lu2014connected}. 
Besides videocameras, \gls{lidar} sensors are often used, as they are the most precise systems to measure range, and robust under almost all lighting and weather conditions with or without glare and~shadows~\cite{li2020lidar}.

Notably, autonomous vehicles will implement computer vision algorithms for detecting and classifying objects and obstacles in the surroundings~\cite{feng2020deep}, including cars, pedestrians, and road signs. 
In this context, more robust scene understanding could be achieved if vehicles exchanged their sensor data via \gls{v2x} communication to other vehicles and/or road infrastructures. This approach permits to extend the perception range beyond the capabilities of onboard instrumentation, a concept usually referred to as \emph{cooperative perception}~\cite{hobert2015enhancements,higuchi2019value}.
However, transmitting large volumes of data may be challenging for current \gls{v2x} communication technologies.
For example, a raw LiDAR frame is, on average, 1 MB~\cite{testolina2022selma}: with a frame rate of 30 fps, it would produce a data rate of around 240 Mbps. For comparison, 3GPP C-V2X and IEEE 802.11p, i.e., the de facto standards for \gls{v2x}, can offer a nominal data rate of only a few tens of Mbps~\cite{zugno2020toward}.
One possible method to solve capacity issues is to compress data before transmission, which in turn introduces additional complications~\cite{nardo2022point}. 
For example, compression may sacrifice accuracy to reduce the file size, with severe implications for the operations that rely on data such as object detection.
Moreover, while compressing data from videocameras is relatively straightforward, there is no accepted standard for compression of \gls{lidar} data.

\subsection{Motivations}
The design of communication algorithms for autonomous driving requires a rigorous process of validation. 
To this aim, using a real testbed is impractical due to limitations in scalability, flexibility, and the high cost of hardware components. 
Theoretical analyses, in turn, often introduce conservative and/or unrealistic assumptions on the system model, and may lead to wrong or misleading conclusions. 
Simulations, in turn, have the advantage to reduce costs and time consumption for validation, and would facilitate the research process. 

To do so, computer simulations need accurate modeling of the different components of the network at all layers. 
Notably, at the application layer, automotive data is required, which comes with at least three main concerns.
First, labeled data are often expensive and time-consuming to generate, or completely unavailable ~\cite{testolina2022selma}.
Second, the dataset shall be available and stored on the simulator machine.
For example, SemanticKITTI~\cite{behley2019semantickitti}, a popular open-source labeled dataset for autonomous driving~research, consists of around 80 GBytes of data, which may consume excessive memory resources.
Third, the dataset shall be accessed, saved, and processed in the RAM of the simulator machine, introducing delays that scale with the size of the dataset.
In particular, data compression requires point-level processing, which may be hard to perform in real time.
For example, geometry-based point cloud (G-PCC)~\cite{graziosi2020overview}, a possible standard for \gls{lidar} compression, can compress only 440k points/s~\cite{nardo2022point} (for comparison, the HDL-64E LiDAR sensor used in SemanticKITTI captures 1.3M points/s). 
Moreover, data compression involves expensive and energy-consuming hardware (particularly high-end GPUs), which may be unavailable on the simulator~machine.

An alternative approach is to simulate the application layer using a statistical model of the automotive data rather than real data.
Therefore, the traffic is simulated based on the sole arrival process of packets (especially the packet size and the inter-arrival time) according to some mathematical process, and 
this approach does not require a dataset to be available and stored on the simulator machine.
Our early results indicate that a single simulation in ns-3, a popular full-stack end-to-end network simulator, of 15 s takes on average around 605 s (10 min) to complete using automotive data from SemanticKITTI, vs. only 20 s using the corresponding statistical model. 

In the literature, statistical methods have been proposed to model, for example, the propagation of the signal~\cite{lecci2021accuracy},
the application (e.g., for web browsing~\cite{cao2004stochastic} or, more recently, \gls{xr} traffic~\cite{lecci2021ns}) or, in the vehicular domain, automotive radar reflections~\cite{buller2013statistical} or multi-sensor data fusions systems~\cite{ahmadi2017statistical}.
However, to date, there is no universal statistical model for automotive sensor data, particularly for LiDAR point clouds. Ideally, such data could be represented as periodic traffic with a fixed frame rate and a constant frame size proportional to the resolution of the sensor. However, in practice, automotive data is often compressed before transmission to reduce the file size, leading to frames of variable size. 
Therefore, automotive data should be modeled as a combination of random variables, after proper fitting and validation via statistical~methods.

\begin{figure*}[t]
    \centering
    \includegraphics[width=0.95\linewidth]{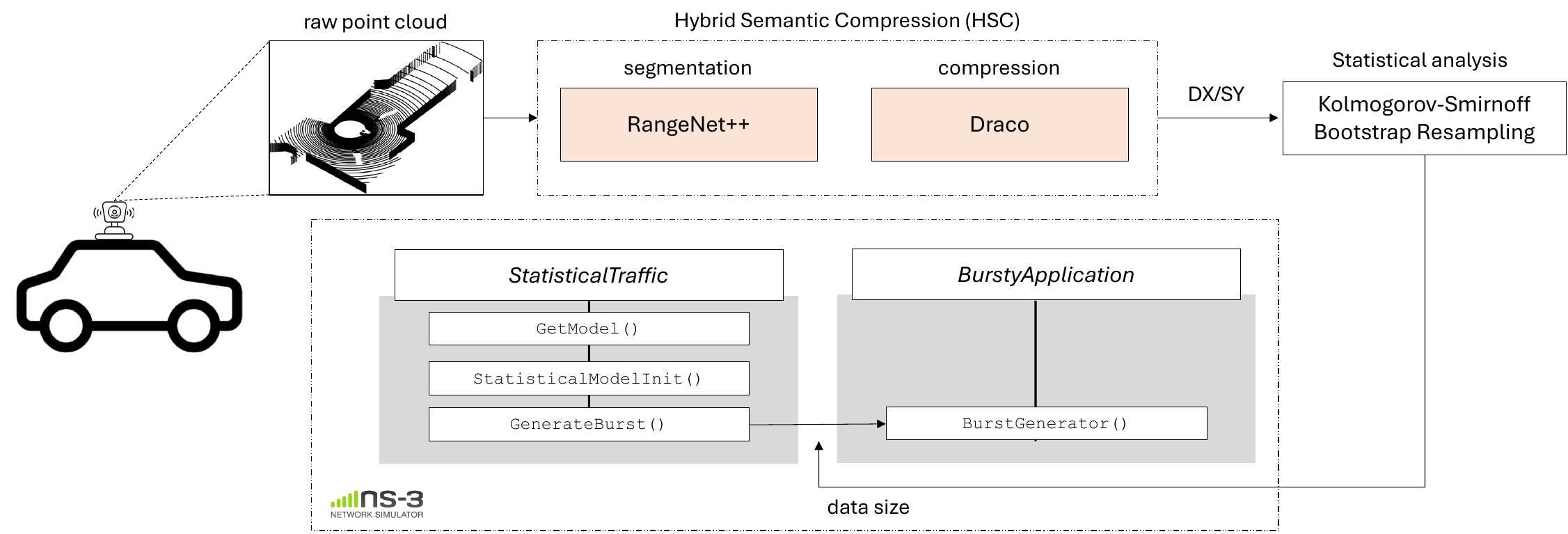}
    \caption{Overview of the system model. An autonomous vehicle generates 3D LiDAR point clouds, which may be compressed via \acrfull{hsc} to reduce the file size. Statistical analysis via \acrfull{ks} and Bootstrap Resampling methods is performed to identify statistical models for the size of 3D LiDAR point clouds. The accuracy of the models is assessed via statistical tests, and based on their impact on network metrics, measured via ns-3 simulations.}
    \label{fig:overview}
\end{figure*}

\subsection{Contributions} 
 \label{sub:contributions}
 Based on the above introduction, the contributions of this paper can be summarized as follows.

 \begin{itemize}
  \item We provide a realistic statistical characterization of automotive data, specifically of the size of LiDAR point clouds. To the best of our knowledge, this is the first model for automotive traffic, and is based on the SemanticKITTI dataset. 
  Given the importance of compression in the automotive scenario, we provide seven different models to characterize raw data and six representative compression configurations. Specifically, data is compressed using the state-of-the-art \gls{hsc} algorithm, first proposed in~\cite{varischio2021hybrid}, which supports different levels of compression to trade off quality against speed. We claim that the availability of statistical models for automotive data brings several advantages compared to using actual data, including faster simulations and processing at the application layer, and no or limited storage of data.
  \item We quantify the accuracy of our statistical models. 
  Specifically, we test different distributions, and identify the corresponding fitting parameters. The measure of accuracy is assessed via a custom statistical test based on the \gls{mle}, \gls{ks}, and Bootstrap Resampling schemes~\cite{wang2011application}. Our results show that the size of the uncompressed data can be accurately modeled according to the tLocationScale distribution, while for the compressed data we select the tLocationScale, Nakagami, Logistic or Gamma distributions according to the level of compression. Interestingly, six of the seven models pass the test.
  \item We further validate the accuracy of our statistical models on representative network metrics via ns-3 simulations. First, we extend the ns-3 code base with new custom methods to generate random variables according to the tLocationScale, Nakagami, and Logicstic distributions, which are not natively available in ns-3. 
  Then, we simulate the transmission of automotive data between two vehicles as a function of their distance. At the application layer, we simulate data transmission at the packet level (raw or compressed) using real data from the SemanticKITTI dataset or based on our statistical traffic models, and compare the corresponding network metrics at the \gls{pdcp} layer. 
  We observe that, even though only six of the seven models passed the test, the statistical approach produces similar, if not the same, results as with real data in terms of end-to-end latency and throughput. This confirms that statistical models can be used to simplify network simulations without compromising the reliability of the results.
\end{itemize}

The rest of the paper is organized as follows.
In Sec.~\ref{sec:related} we present some related works. In Sec.~\ref{sec:sys_model} we briefly describe our system model, specifically the HSC pipeline for data compression. In Sec.~\ref{sec:traffic_analysis} we formalize our statistical analysis to define accurate models for automotive data, and assess the accuracy of those models. In Sec.~\ref{sec:ns3} we describe the ns-3 implementation of the random distributions used by the models. In Sec.~\ref{sec:performance_evaluation} we present simulation results to evaluate the impact of the models on the network under several metrics. 
Finally, conclusions and suggestions for future work are discussed in Sec.~\ref{sec:conclusions}.

\section{Related Work}
\label{sec:related}
\gls{v2x} communication, both among vehicles and/or with the network, is a key enabler of the autonomous driving revolution.
However, it comes with several challenges, due to the large size of the data~\cite{choi2016millimeter}, 
the highly dynamic and heterogeneous road environment, as well as the stringent network requirements of autonomous driving~\cite{3gpp.22.186}.
In this sense, a complete characterization of the traffic generated by autonomous vehicles is essential to design the appropriate \gls{v2x} communication technologies and protocols. 
Furthermore, as network functions become increasingly virtualized, traffic modeling is critical for proper network dimensioning~\cite{huo2022sdn}.
Additionally, (generative) traffic models play a crucial role in network simulations, significantly reducing both the cost and time required for implementing and testing new solutions and architectures~\cite{zerwas2021network}. 
For these reasons, modeling the network traffic has been a central research topic for the last decades.

Authors in~\cite{navarro2020survey} presented an extensive list of traffic models for \gls{5g} applications, though with limited references to the autonomous driving scenario.
Several works modeled the data stream for \gls{iot} applications. 
Specifically, the authors in \cite{bulashenko2020new} modeled the aggregated \gls{m2m} traffic and the corresponding message delivery delay in a \gls{5g} network, while papers \cite{osterbo2017state} and \cite{sansoni2018comparison} focused on a single traffic source, deriving stochastic models based on Markov Chains.
In addition, the survey in~\cite{lao2020survey} reported an overview of traffic models for \gls{p2p} communication in \gls{iot} blockchain networks.
The traffic generated by video streaming has also been extensively investigated~\cite{tanwir2013survey}.
For example, Kalbkhani \emph{et al.}~\cite{kalbkhani2017adaptive} proposed to use non-linear autoregressive models to predict the future frame size in video traffic.
Besides classic video applications, the research community is working on the characterization of interactive video traffic for \gls{xr} applications, e.g., in~\cite{lecci2022temporal,chiariotti2023temporal}.
In addition to the more classical models based on state machine and autoregression techniques~\cite{osterbo2017state,sansoni2018comparison,grigoreva2017coupled}, more recently Machine and Deep Learning techniques were employed to model complex traffic patterns.
For instance, Nie \emph{et al.}~\cite{nie2018network} combined a deep-belief network with a compressed-sensing approach to predict the fast- and slow-varying traffic components  in a wireless mesh network.
Other works such as \cite{zhuo2017long,zhang2021lntp} introduced \gls{lstm} models to capture and forecast network traffic statistics.

For the specific case of autonomous driving, only a few studies have specifically addressed the characterization of the network traffic.
For example, Choi \emph{et al.}~\cite{choi2016millimeter} characterized the data rates of different automotive sensors, using information obtained from the datasheets of commercial products and conversations with industrial partners; for LiDAR (videocamera) sensors, the resulting data rate was measured between 10 and 100 (100 and 700) Mbps depending of the resolution.
Grigoreva \emph{et al.}~\cite{grigoreva2017coupled} further introduced a machine-type communication traffic model tailored to automotive applications, incorporating spatial and temporal correlations.
Wang  \emph{et al.}~\cite{wang2020real} measured the mean frame size, frame rate, and delay of sensor data based on an experimental demo system, providing additional empirical insights.
Similarly, the data rate of the commercial Velodyne HDL-64E LiDAR sensor was characterized in~\cite{chen2019low}.
However, these models have been obtained for specific types of sensors, and only model the data rate of the sensors.
A comprehensive, closed-form, and statistical characterization of automotive data is still an open challenge, which motivates the research presented in this paper.

\section{System Model}\label{sec:sys_model}
In this section we present our system model, also illustrated in Fig.~\ref{fig:overview}, specifically the automotive data (Sec.~\ref{sub:automotive_data}), the \gls{hsc} compression pipeline (Sec.~\ref{sec:hsc}), and the selection process to identify the types of data to be analyzed in this paper (Sec.~\ref{sec:compression_model_refinement}),

\subsection{Automotive Data} 
\label{sub:automotive_data}
Autonomous vehicles rely on a combination of multiple sensors to perceive their surroundings. In addition to videocameras, LiDAR sensors are often used, given their ability to provide accurate distance measurements in different lighting and weather conditions.
Therefore, in this paper we propose a statistical characterization of the size of automotive data focusing on LiDAR sensors, and based on SemanticKITTI~\cite{behley2019semantickitti}, a large-scale, high-resolution dataset designed for semantic segmentation tasks in autonomous driving research.
It extends the popular KITTI Vision Benchmark Suite by providing dense, point-wise annotations for LiDAR data collected using a Velodyne HDL-64E sensor. The dataset consists of 22 sequences, for a total of 43\,552 LiDAR scans, captured in diverse environments such as urban streets, rural areas, and highways. Each point in the 3D point cloud is annotated based on 28 semantic classes, including road, building, vegetation, and dynamic objects like cars, pedestrians, and~cyclists.

\subsection{Hybrid Semantic Compression (HSC)}\label{sec:hsc}
Data transmission is challenging, given the large size of LiDAR point clouds. 
For SemanticKITTI, each raw LiDAR acquisition generates a point cloud of around 120\,000 points, with an average file size of around 3\,200 KB. 
Therefore, data should be processed, e.g., compressed, before transmission, to reduce the file size and so the link~overload.

In \cite{varischio2021hybrid}, we proposed a compression pipeline for LiDAR data called \gls{hsc}, which exploits the semantic understanding of the 3D scene to reduce the size of point clouds.
The HSC pipeline consists of two modules: a Deep Learning semantic module (based on RangeNet++~\cite{RangeNetpp}) to select safety-critical points, and a compression module (based on Google Draco~\cite{Draco}) to compress the resulting point cloud at fast speed.
Specifically, we consider three \glspl{sl}: SL $=0$, representing raw data before compression; SL  $=1$, where we remove points from the point cloud relative to the road and the background; and SL $=2$, where points from buildings, vegetation, and traffic signs are also removed.
For compression, Draco relies on two parameters: the \gls{qp} $\in\{1,\dots,14\}\bigcup\{0\}$, i.e., the number of bits used for quantizing input values, with \gls{qp} $=0$ indicating no quantization; and the \gls{cl} $\in\{0,\dots,10\}$, which trades off compression accuracy over efficiency (measured in terms of the encoding and decoding time). More precisely, higher values of \gls{cl} achieve better compression, at the cost of a slower encoding or decoding.

\subsection{Data Selection Process}\label{sec:compression_model_refinement}
Overall, HSC results in a rich set of (almost 500) possible compression configurations, offering a fine-grained control over the compression performance.
However, for simplicity, in this work we decided to consider only a selection of seven representative compression configurations.
Specifically, we inspected the rendered point clouds after Draco compression, and grouped similar configurations based on (i) the impact of compression on the quality of data, and (ii) the average encoding and decoding time.
Assuming that, for lossy compression, the main source of distortion is quantization, we investigate the effect of \gls{qp} on the compressed point clouds, and study the encoding and decoding time as a function of \gls{cl}.

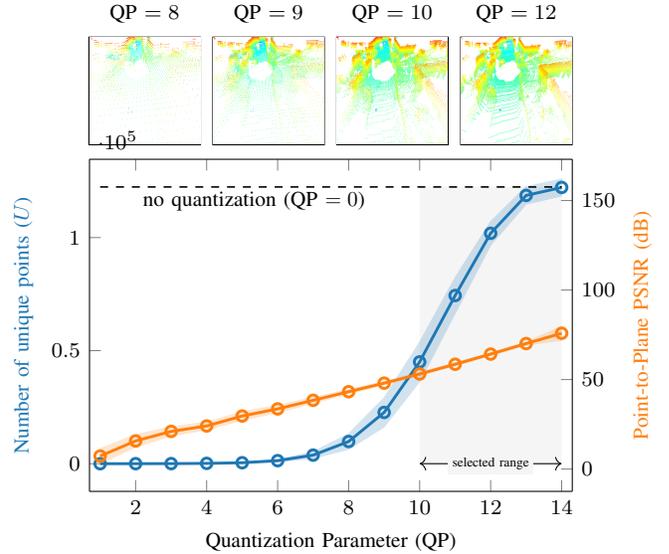
\begin{figure}[t!]
    \centering
\pgfplotsset{
tick label style={font=\footnotesize},
label style={font=\footnotesize},
legend  style={font=\footnotesize}
}\begin{tikzpicture}

\definecolor{darkgray176}{RGB}{176,176,176}
\definecolor{darkorange}{RGB}{255,127,14}
\definecolor{steelblue}{RGB}{31,119,180}
\definecolor{whitesmoke}{RGB}{245,245,245}

\usepgfplotslibrary{fillbetween}

\begin{axis}[name=r1,
            height=3cm,
            width=3.05cm,
            xshift=0.178cm,
            title={\footnotesize QP $=8$},
            title style={at={(axis description cs:0.5,1.25)},anchor=north},
            xmin=-0.5, xmax=1398.5,
            ymin=-0.5, ymax=751.5,
            y dir=reverse,
            ticks=none,
            ,
            axis line style = thick
            ]
    \addplot graphics [includegraphics cmd=\pgfimage,xmin=-0.5, xmax=1398.5, ymin=751.5, ymax=-0.5] {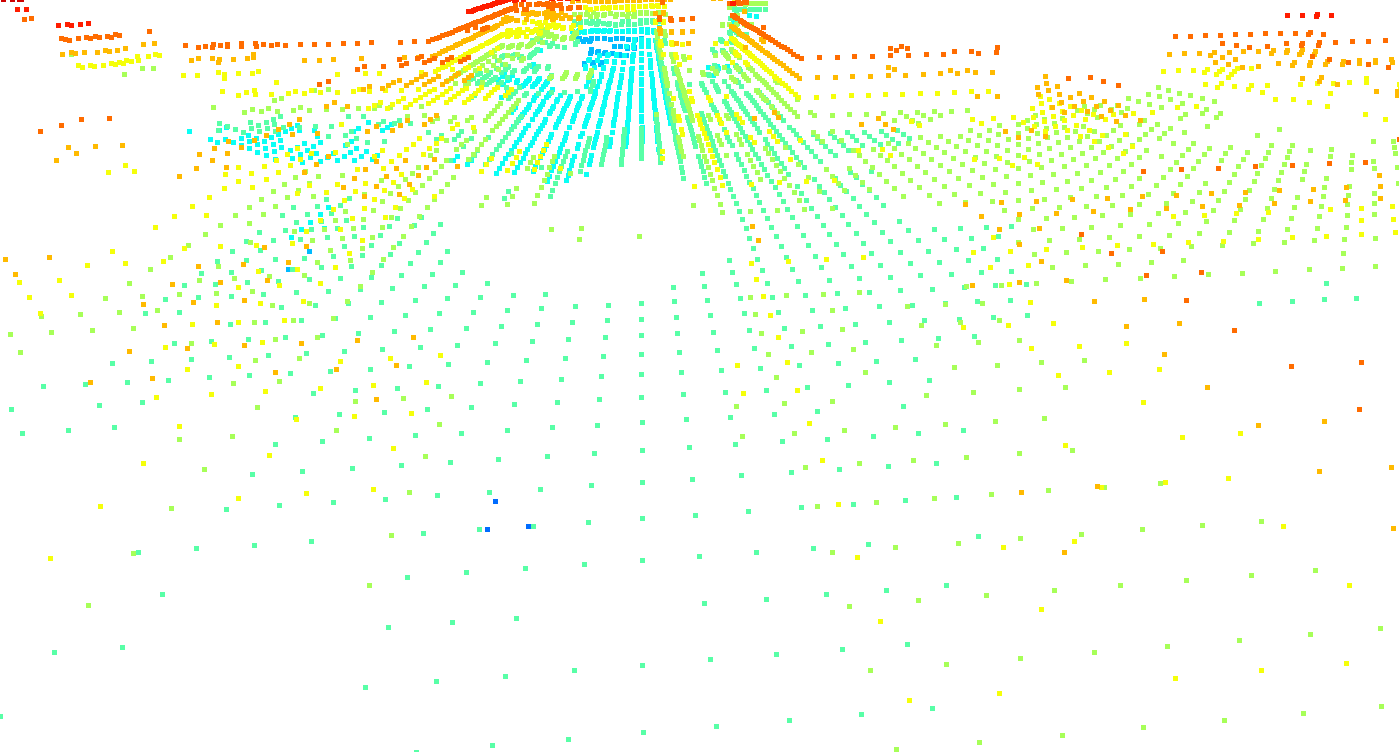};
\end{axis}


\begin{axis}[name=r2,
            at={(r1.east)},
            anchor=west,
            height=3cm,
            title={\footnotesize QP $=9$},
            title style={at={(axis description cs:0.5,1.25)},anchor=north},
            width=3.05cm,
            xshift=0.178cm,
            xmin=-0.5, xmax=1398.5,
            ymin=-0.5, ymax=751.5,
            y dir=reverse,
            ticks=none,
            axis line style = thick
            ]
    \addplot graphics [includegraphics cmd=\pgfimage,xmin=-0.5, xmax=1398.5, ymin=751.5, ymax=-0.5] {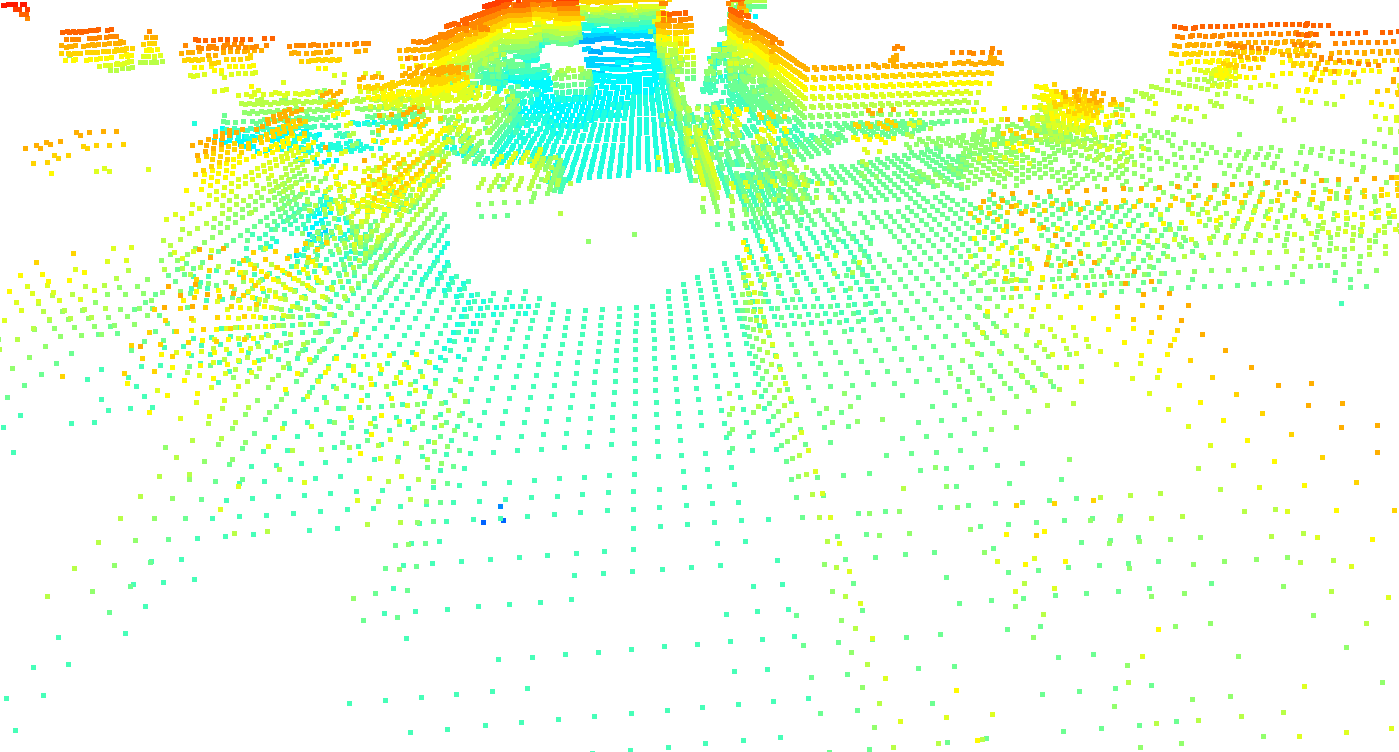};
\end{axis}

\begin{axis}[name=r3,
            at={(r2.east)},
            anchor=west,
            height=3cm,
            title={\footnotesize QP $=10$},
            title style={at={(axis description cs:0.5,1.25)},anchor=north},
            width=3.05cm,
            xshift=0.178cm,
            xmin=-0.5, xmax=1398.5,
            ymin=-0.5, ymax=751.5,
            y dir=reverse,
            ticks=none,
            axis line style = thick
            ]
    \addplot graphics [includegraphics cmd=\pgfimage,xmin=-0.5, xmax=1398.5, ymin=751.5, ymax=-0.5] {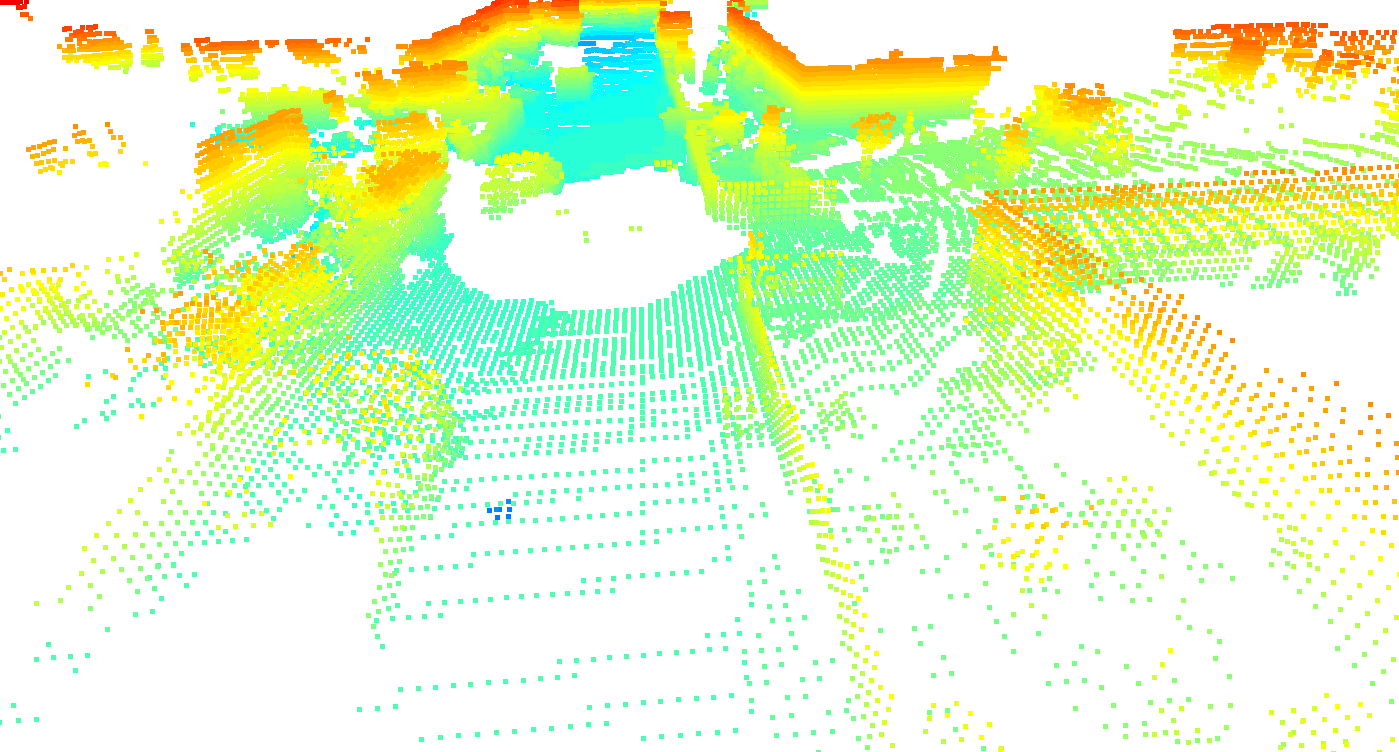};
\end{axis}

\begin{axis}[name=r4,
            at={(r3.east)},
            anchor=west,
            height=3cm,
            title={\footnotesize QP $=12$},
            title style={at={(axis description cs:0.5,1.25)},anchor=north},
            width=3.05cm,
            xshift=0.18cm,
            xmin=-0.5, xmax=1398.5,
            ymin=-0.5, ymax=751.5,
            y dir=reverse,
            ticks=none,
            axis line style = thick
            ]
    \addplot graphics [includegraphics cmd=\pgfimage,xmin=-0.5, xmax=1398.5, ymin=751.5, ymax=-0.5] {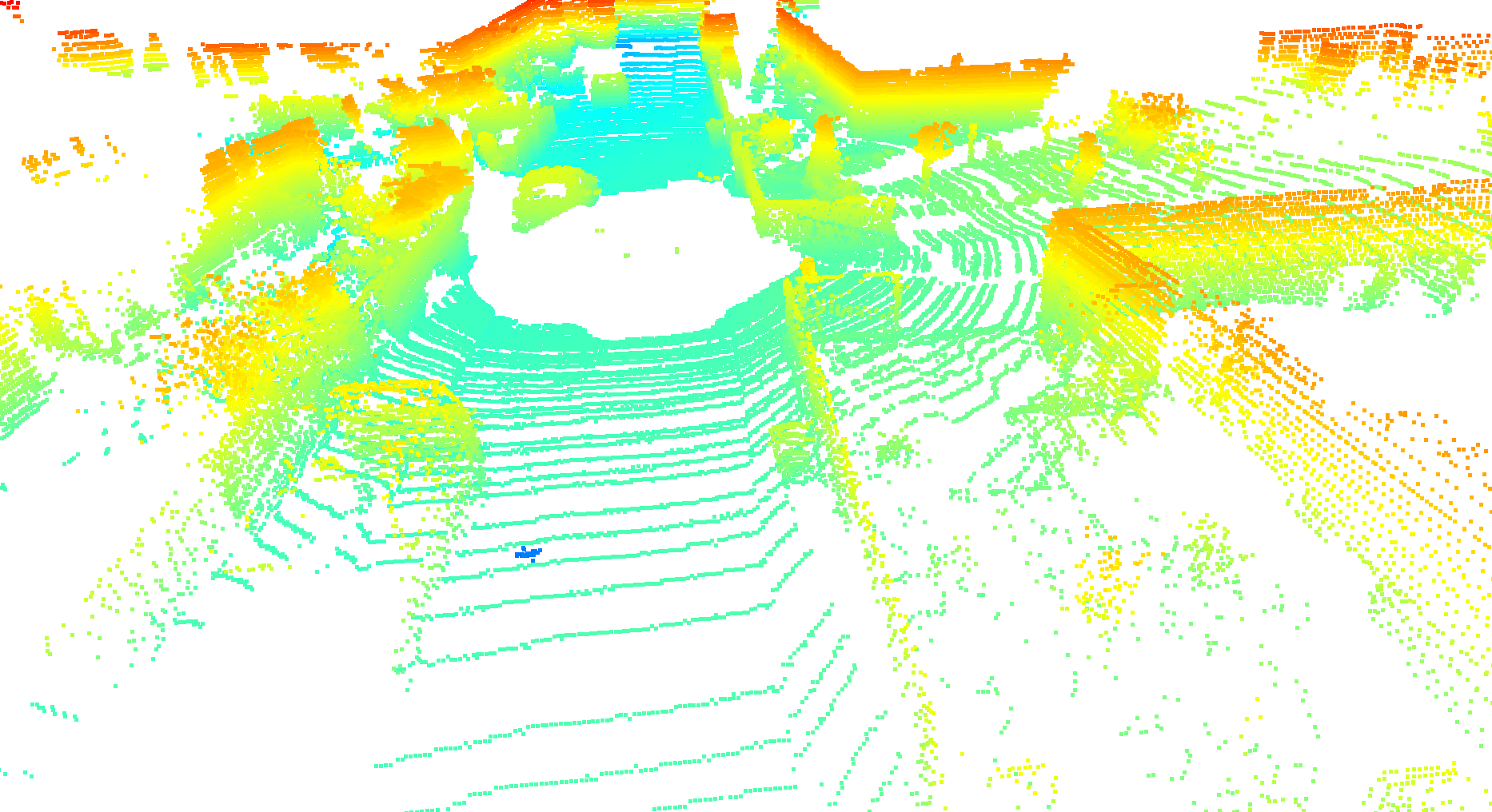};
\end{axis}
\begin{axis}[
    name=main,
    at={(r1.south west)},
    anchor=north west,
    yshift=-0.2cm,
    width=8.cm,
    xlabel={Quantization Parameter (QP)},
    ylabel={Number of unique points ($U$)},
    ylabel style ={steelblue},
    axis y line*=left,
    height=6cm,
    xmin=0.7,
    xmax=14.3,
    ytick align=inside,
    xtick align=inside,
    minor xtick={},
    ]

    \addplot[color=steelblue,mark=o , mark size=2pt, line width=1.1pt]
        table[x=qp, y=n_points, col sep=space] {Figures/renders/n_points.txt}; \label{plot_one}
    
    \path [draw=steelblue, fill=steelblue, opacity=0.2]
    (axis cs:1,2.72818233825688)
    --(axis cs:1,4.31727220719767)
    --(axis cs:2,15.1592782016723)
    --(axis cs:3,51.0100175365553)
    --(axis cs:4,201.041176934288)
    --(axis cs:5,591.763226666287)
    --(axis cs:6,1863.58241289594)
    --(axis cs:7,5229.89080791636)
    --(axis cs:8,13209.1851644237)
    --(axis cs:9,29106.6834190169)
    --(axis cs:10,53727.2287167938)
    --(axis cs:11,82282.0251036159)
    --(axis cs:12,107112.629966863)
    --(axis cs:13,122484.800622746)
    --(axis cs:14,125935.424433649)
    --(axis cs:14,118419.339202714)
    --(axis cs:14,118419.339202714)
    --(axis cs:13,114809.381195436)
    --(axis cs:12,96686.7245785911)
    --(axis cs:11,66466.1476236569)
    --(axis cs:10,36327.8440104789)
    --(axis cs:9,16252.6983991649)
    --(axis cs:8,6396.29665375807)
    --(axis cs:7,2348.44555572)
    --(axis cs:6,828.290314376786)
    --(axis cs:5,265.782227879168)
    --(axis cs:4,83.2497321566216)
    --(axis cs:3,22.7808915543538)
    --(axis cs:2,7.76799452560047)
    --(axis cs:1,2.72818233825688)
    --cycle;

    \addplot [name=noquant, semithick, black, dashed]
    table {%
    1. 122393.559090909
    14. 122393.559090909
    };

    \draw (axis cs:2,114000.559090909) node[
      scale=.8,
      anchor=base west,
      text=black,
      rotate=0.0
    ]{no quantization (\gls{qp} $=0$)};
    
    \draw[<->,draw=black] (axis cs:10,10) -- (axis cs:14,10);
    \draw (axis cs:12,100) node[
      scale=0.5,
      fill=whitesmoke,
      draw=none,
      text=black,
    ]{selected range};
    
    \addplot [name path=left,draw=none,thick,mark=none,smooth] coordinates {
                (10, 3) (10, 122394) };
    \addplot [name path=right,draw=none,thick,mark=none,smooth] coordinates {
                (14, 3) (14, 122394) };
    \addplot [whitesmoke] fill between[
        of=left and right,
        ];  
\end{axis}

\begin{axis}[
    name=main2,
    at={(main.north west)},
    anchor=north west,
    yshift=-0.2cm,
    axis y line*=right,
    axis x line=none,
    ylabel={Point-to-Plane PSNR (dB)},
    ylabel style ={darkorange},
    height=6cm,
    width=8cm,
    xmin=0.7,
    xmax=14.3,
    ytick align=inside,
    minor xtick={},
]

\addplot[color=darkorange, mark=o, mark size=2pt, line width=1.1pt]
    table[x=qp, y=psnr, col sep=space] {Figures/renders/psnr.txt}; \label{plot_two}

\path [draw=darkorange, fill=darkorange, opacity=0.2]
(axis cs:1,3.08840120077685)
--(axis cs:1,11.4430659576666)
--(axis cs:2,19.2451046024534)
--(axis cs:3,23.540605236728)
--(axis cs:4,26.2130071187376)
--(axis cs:5,31.9715282127443)
--(axis cs:6,35.6915523749665)
--(axis cs:7,40.0305405321089)
--(axis cs:8,44.3835458525018)
--(axis cs:9,48.9358724727015)
--(axis cs:10,53.9869202384299)
--(axis cs:11,59.3822456847887)
--(axis cs:12,65.076826055659)
--(axis cs:13,70.9801848834951)
--(axis cs:14,79.7222267706139)
--(axis cs:14,71.9924905544667)
--(axis cs:14,71.9924905544667)
--(axis cs:13,69.2372888276643)
--(axis cs:12,63.3654380105947)
--(axis cs:11,57.7014248569304)
--(axis cs:10,52.2384507874886)
--(axis cs:9,47.0498520432253)
--(axis cs:8,42.0421391065889)
--(axis cs:7,36.7843387555487)
--(axis cs:6,31.5340041616951)
--(axis cs:5,27.3015872231994)
--(axis cs:4,21.9318510526812)
--(axis cs:3,18.5313913019903)
--(axis cs:2,12.1745144513624)
--(axis cs:1,3.08840120077685)
--cycle;

\addplot [semithick, darkorange, dashed, draw=none]
table {%
1 157.594591098694
14. 157.594591098694
};
\addplot [name path=left,draw=none,thick,mark=none,smooth, draw=none] coordinates {
            (10, 3) (10, 157.594591098694) };

\end{axis}

\end{tikzpicture}






    \caption{Number of unique points $U$ and p2p-PSNR after compression vs. \gls{qp}. Above, random point clouds from SemanticKITTI after compression for QP $\in\{8,9,10,12\}$.}
    \label{fig:QP_refinement}
\end{figure}

\subsubsection{Compression quality analysis}
The semantic understanding of data, e.g., detection and classification of objects in the environment, is a key component for the development of reliable autonomous vehicles~\cite{Wang_2019_CVPR,8374608,Zhang_2021_ICCV,Zhang_2020_CVPR,squeezeSeg}.
However, the semantic performance may rapidly deteriorate due to lossy compression of the input~\cite{effectJpegClassification, s22031104, jo2021impact}.

Thus, we analyze the impact of HSC Draco compression on the quality of LiDAR point clouds. 
Specifically, we set \gls{cl} $=7$, and vary \gls{qp} from 0 to 14 as per Draco's specifications. Then, we compute:
\begin{itemize}
    \item {The number $U$ of unique points in the compressed point cloud}: Draco preserves the number of points between input and output but, as a result of the compression process, it condenses multiple points into a single one, reducing the number of unique points in the compressed point cloud.
    \item {The \gls{p2pPSNR}} \cite{p2pPSNR} to quantify the geometric distortion of point cloud compression based on the \gls{psnr}.
\end{itemize}
Both metrics are computed and averaged over a sample of 550 LiDAR point clouds, uniformly selected from the SemanticKITTI dataset.
In Fig.~\ref{fig:QP_refinement} we plot $U$ and \gls{p2pPSNR} as a function of \gls{qp}, and also report random point cloud samples from SemanticKITTI for QP $\in\{8,9,10,12\}$ for a qualitative evaluation. 
We observe that, for \gls{qp} $\leq10$, compression is overly aggressive as the visual quality of the point clouds is severely degraded.
Furthermore, at QP $=10$, we empirically notice a change in the slope of the \gls{p2pPSNR} curve from approximately $5$~dB/\gls{qp} to $6.25$~dB/\gls{qp}, as well as a decrease of the number of unique post-compression points by approximately $70$\%.
Hence, based on these observations, we selected \gls{qp} values in the range $\{10, 11, 12, 13, 14,\}\bigcup\{0\}$.
Notice that this analysis is consistent with the algorithm description in the Draco repository,\footnote{Draco code repository: \hyperlink{https://github.com/google/draco}{https://github.com/google/draco}} where it is stated that \emph{``most projects can set quantization values of about $11$ without any noticeable difference in quality''}.

\begin{figure}[t!]
    \centering
\definecolor{darkgray176}{RGB}{176,176,176}
\definecolor{darkorange}{RGB}{255,127,14}
\definecolor{steelblue}{RGB}{31,119,180}
\definecolor{whitesmoke}{RGB}{245,245,245}
\pgfplotsset{
tick label style={font=\footnotesize},
label style={font=\footnotesize},
legend  style={font=\footnotesize}
}
\begin{tikzpicture}
\def\hratio{0.18}
\def\wratio{0.45}
\def\xshift{0.9}
\def\colorpalette{Dark 2}
\def\gridopacity{20}
\def\qqplotBisectorColor{red}

\pgfplotsset{cycle list/Reds-6}

\begin{axis}[
    name=filesize,
    ylabel={Compressed file size [KB]},
    minor xtick={},
    xmin=0,
    xmax=10,
    grid=both,
    grid style={line width=.5pt, draw=gray!\gridopacity},
    enlargelimits=0.1,
    width=\wratio\textwidth,height=\hratio\textheight,
    ymode=log,
    xticklabels=\empty,
            legend cell align={left},
legend columns=3,
legend style={
  fill opacity=0.8,
  draw opacity=1,
  text opacity=1,
  at={(0.5,1.5)},
  anchor=north,},
]
\foreach \column in {10,11,12,13,14,0}{
  \addplot+[ycomb,line width=1.1pt, mark=o] table[x=index,y expr=\thisrow{\column}/1000, col sep=comma] {Figures/Quantization_Plots/data/quant_filesize.csv};
  }

    \legend {QP $=10$,QP $=11$,QP $=12$,QP $=13$,QP $=14$,QP $=0$ };
\end{axis}
\end{tikzpicture}

\begin{tikzpicture}
\def\hratio{0.18}
\def\wratio{0.45}
\def\xshift{0.9}
\def\colorpalette{Dark 2}
\def\gridopacity{20}
\def\qqplotBisectorColor{red}

\pgfplotsset{cycle list/Reds-6}

\begin{axis}[
    name=enctime,
    at={(filesize.north east)},
    anchor=north west,
    ylabel={Encoding time [ms]},
    minor xtick={},
    xmin=0,
    xmax=10,
    grid=both,
    xticklabels=\empty,
    grid style={line width=.5pt, draw=gray!\gridopacity},
    enlargelimits=0.1,
    width=\wratio\textwidth,height=\hratio\textheight,
]

\foreach \column in {10,11,12,13,14,0}{
  \addplot+[ycomb,line width=1.1pt, mark=o] table[x=index,y=\column,col sep=comma] {Figures/Quantization_Plots/data/quant_enctime.csv};
  }

\end{axis}
\end{tikzpicture}

\begin{tikzpicture}
\def\hratio{0.18}
\def\wratio{0.45}
\def\xshift{0.9}
\def\colorpalette{Dark 2}
\def\gridopacity{20}
\def\qqplotBisectorColor{red}

\pgfplotsset{cycle list/Reds-6}

\begin{axis}[
    name=dectime,
    at={(enctime.north east)},
    anchor=north west,
    xshift=\xshift cm,
    xlabel={Compression Level (CL)},
    ylabel={Decoding time [ms]},
    minor xtick={},
    xmin=0,
    xmax=10,
    grid=both,
    grid style={line width=.5pt, draw=gray!\gridopacity},
    enlargelimits=0.1,
    width=\wratio\textwidth,height=\hratio\textheight,
]

\pgfplotsinvokeforeach{10,11,12,13,14,0}{
  \addplot+[ycomb,line width=1.1pt, mark=o, mark size=2pt] table[x=index,y=#1,col sep=comma] {Figures/Quantization_Plots/data/quant_dectime.csv};
  }

\end{axis}
\end{tikzpicture}
    \caption{Average file size and encoding/decoding time vs. \gls{cl} $\in \{0,1,\dots, 10\}$, and for \gls{qp} $\in\{0,10,11,12,13,14\}$.}
    \label{fig:cl_analysis}
\end{figure}
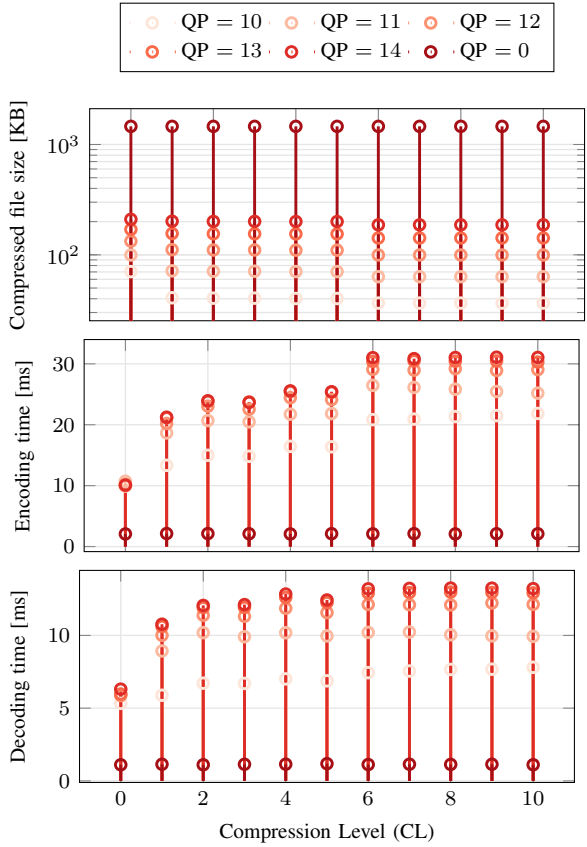

\subsubsection{Compression time analysis}
Based on the values of \gls{qp} previously selected, we now calculate the file size, encoding time, and decoding time for compression vs. \gls{cl} $\in\{0,1,\dots, 10\}$.\footnote{Note that the parameter combination \{\gls{qp}$=0 $, \gls{cl}$=0$\} does not disable Draco, thus it does not correspond to raw data. In fact, while the raw data file from SemanticKITTI has the extension .ply, the output file after Draco is applied has extension .drc, regardless of the compression mode, which is a more efficient data format representation. This also explains why the encoding and decoding times are not zero in this case.}
In Fig.~\ref{fig:cl_analysis} we observe that, for a fixed value of \gls{qp}, the file size does not significantly change as a function of \gls{cl}. 
Conversely, we see that the encoding and decoding times increase as CL increases, especially for \gls{cl} $\geq5$, with marginal effects on the file size. 
Moreover, the curves almost overlap for QP $\in\{12,13,14\}$. 
For this reason, we choose the corner case QP $=14$ as a cluster representative, and fix \gls{cl} $=5$ as it represents a sweet spot between file size and encoding/decoding time.

\subsubsection{Final selection} 
\label{ssub:final_selection}

In summary, we select a subset of Draco configurations that are deemed {qualitatively acceptable and most relevant} for \gls{lidar} automotive data, specifically \gls{qp} $\in\{10, 11,14\}$ and $\text{\gls{cl}}=5$.
Combined with the three \glspl{sl} of \gls{hsc}, we define $9$ models, identified as \textbf{DX/SY}, where X is the QP, and Y is for the SL~$\in\{0,1,2\}$. 
We use the value X $=0$ to represent the configuration where Draco is disabled. For example, D0/S0 identifies raw data, and D0/S1 represents raw data with SL $=1$.
Finally, to further reduce the number of combinations to analyze, we neglect $\text{\gls{qp}}=10$, and consider only $\text{SL}=0$ for $\text{\gls{qp}}=11$ (D11/S0), which represents a pure Draco compression with no \gls{hsc} semantic module.
In this way, the cardinality of the model space is effectively reduced to seven combinations:
D0/S0, D0/S1, D0/S2, D11/S0, D14/S0, D14/S1, and D14/S2.

In the next section, we will provide statistically representative traffic models to characterize the size of LiDAR data based on the seven HSC configurations selected above.

\section{Statistical Models for Automotive Data}\label{sec:traffic_analysis}
Assuming a constant generation interval for LiDAR data (typically 10 or 30 fps), the size of the point clouds to be transmitted is the most relevant element to be considered when characterizing the source traffic.
Thus, we consider the SemanticKITTI dataset~\cite{behley2019semantickitti}, and evaluate the \glspl{cdf} of the size of LiDAR point clouds for each of the compression configurations defined in Sec.~\ref{sec:compression_model_refinement}.
In Sec.~\ref{ssub:statistical_test} we describe our statistical method, while validation results are reported in Sec.~\ref{ssub:statistical-results}.

\emph{Notation}: We denote vectors as bold, lower case letters.

\subsection{Statistical Method} 
\label{ssub:statistical_test}

For each compression configuration, we consider a diverse set $\mathbb{P}$ of theoretical \gls{cdf} families as potential candidates for the unknown target \gls{cdf} $F$ that represents the size of such data.
A hypothetical CDF $F_i$, $i\in\{1,2,\dots,|\mathbb{P}|\}$, is tested against the empirical \gls{cdf} (e\gls{cdf}) $\tilde{F}^{(N)}$ derived from a random sample of $N$ observations from $F$. 
This is a well-known procedure called hypothesis testing. 
Specifically, 
in this work we perform a \gls{ks} test \cite{massey1951kolmogorov} to compare $F_i$ and $\tilde{F}^{(N)}$ based on the KS statistic $D^{(N)}$, which quantifies the ``distance'' between the hypothetical distribution and the distribution of the observed data, i.e.,
\begin{equation}\label{eq:kolmogorov_statistic}
    D^{(N)}(i) = \max_x{|\tilde{F}^{(N)}(x)-F_i(x)|}.
\end{equation}
In general, for continuous \glspl{cdf} and assuming that the two samples $\tilde{F}^{(N)}(x)$ and $F_i(x)$ come from the same distribution (null hypothesis), $\tilde{F}^{(N)} \to F_i$ as $N\to\infty$ (strong law of large numbers), thereby $D^{(N)}(i)$ converges to zero almost surely. 
Moreover, under the same hypothesis, the limiting distribution of $\sqrt{N}D^{(N)}(i)$ converges to a Kolmogorov distribution, independent of $F_i$. As a result, under the null hypothesis and for $N$ sufficiently large, Eq.~\eqref{eq:kolmogorov_statistic} has a known distribution, and well-known critical values.
Notice that the interpretation of the KS statistic centers around the p-value and a predefined significance level $\alpha$: if the p-value is less than $\alpha$, the test concludes that the null hypothesis is unlikely to be true, and that it shall be rejected, i.e., the test is not passed for $F_i$.
For ease of notation, we will write $\tilde{F}$ and $D$ instead of $\tilde{F}^{(N)}$ and $D^{(N)}$, respectively, where the dependence on $N$ is implicit, unless stated otherwise.

\begin{table*}[t]
    \centering
    \caption{P-values $\pi_{i,m}^*$ computed according to Eq.~\eqref{eq:pvalues}, where $i\in\mathbb{P}$ represents the row index, while $m$ is the column index and identifies the HSC compression configuration under test. For each configuration, we represent in bold the p-values that pass the test, i.e., $\pi_{i,m}^*\in\mathcal{I}_m$, and highlight in gray the entries selected as best fits, that is $i_m^\star$.}
    \begin{tabular}{llllllll}
    \toprule
    $\mathbb{P}$ & D0/S0 & D0/S1 & D0/S2 & D11/S0 & D14/S0 & D14/S1 & D14/S2 \\
    \midrule
        BirnbaumSaunders & 0 & 0.0 & \textbf{0.061} & \textbf{0.013} & 0.0 & 0.0 & 0.0 \\
        ExtremeValue & 0 & 0.0 & 0.0 & 0.0 & 0.0 & 0.003 & 0.0 \\
        Gamma & 0 & 0.0 & \cellcolor{black!15}\textbf{0.375} & \textbf{0.114} & 0.0 & 0.0 & \cellcolor{black!15}\textbf{0.012} \\
        GeneralizedExtremeValue & 0 & 0.003 & \textbf{0.039} & 0.006 & 0.0 & 0.0 & 0.0 \\
        HalfNormal & 0 & 0.0 & 0.003 & 0.0 & 0.0 & 0.0 & 0.0 \\
        InverseGaussian & 0 & 0.0 & 0.002 & \textbf{0.018} & 0.0 & 0.0 & 0.0 \\
        Logistic & 0 & 0.009 & 0.0 & 0.0 & \cellcolor{black!15}\textbf{0.036} & \textbf{0.024} & 0.0 \\
        Loglogistic & 0 & 0.0 & \textbf{0.018} & \textbf{0.037} & 0.01 & 0.0 & 0.0 \\
        Lognormal & 0 & 0.0 & \textbf{0.065} & \textbf{0.02} & 0.0 & 0.0 & 0.0 \\
        Nakagami & 0 & 0.001 & 0.0 & \cellcolor{black!15}\textbf{0.128} & 0.0 & 0.0 & 0.004 \\
        Normal & 0 & \cellcolor{black!15}\textbf{0.109} & 0.0 & \textbf{0.012} & 0.0 & 0.0 & 0.0 \\
        Poisson & 0 & 0.0 & 0.0 & 0.0 & 0.007 & 0.0 & 0.0 \\
        Rayleigh & 0 & 0.0 & 0.0 & 0.0 & 0.0 & 0.0 & 0.0 \\
        tLocationScale & \cellcolor{black!15}0 & \textbf{0.029} & 0.0 & \textbf{0.014} & \textbf{0.026} & \cellcolor{black!15}\textbf{0.026} & 0.0 \\
        Weibull & 0 & \textbf{0.044} & \textbf{0.15} & 0.0 & 0.003 & 0.0 & 0.01 \\
    \bottomrule
\end{tabular}

    \label{tab:pvalues}
\end{table*}

We denote the set of 3D LiDAR point clouds in the SemanticKITTI dataset as 
\begin{equation}
 \mathcal{D}=\{ \bm{p}_j: \bm{p}_j\in\mathbb{R}^{3\times n_j}, j=1,\dots,N \},
\end{equation}
where $n_j$ is the number of points in the point cloud $\bm{p}_j\in\mathcal{D}$.
We represent each \gls{hsc} compression configuration (denoted by DX/SY) by a function $m: \mathcal{D}\to\hat{\mathcal{D}}$ that receives as input a point cloud $\bm{p}\in \mathcal{D}$, and gives as output its compressed version $\hat{\bm{p}}\in\hat{\mathcal{D}}$. 
Then, we denote as $B_m$ the set of sizes of the point clouds in $\hat{\mathcal{D}}$, compressed using configuration $m$, i.e.,
\begin{align}
    B_m(\hat{\mathcal{D}})=\{ b_j&\in\mathbb{R}:\ b_j=g(\hat{\bm{p}}_j),\\
    &\hat{\bm{p}}_j=m(\bm{p}_j)\in \hat{\mathcal{D}},\ j=1,\dots,N\},
\end{align}
where $g(\bm{p})$ corresponds to the number of bits used to encode $\bm{p}$, i.e., the size of $\bm{p}$.
The goal of our analysis is to find, for each compression configuration $m$, the class of functions $\mathcal{P}_{i}=\{F_{i}(\,\cdot;\bm{\theta}):\bm{\theta}\in\Theta_i\}\in\mathbb{P}$, in the presence of unknown parameters $\bm{\theta}\in\Theta_i$, and the corresponding CDF $F_{i}(\cdot;\bm{\theta}_{i})$, $i\in\{1,2,\dots,|\mathbb{P}|\}$ (our hypothesis, where $\bm{\theta}_i$ is a specific realization of $\bm{\theta}$), that best fit $B_m(\hat{\mathcal{D}})$ (sample of observation) according to some evaluation metric $h_m:\mathcal{P}_i\to\mathbb{R}$. 
In other words, our goal is to perform a goodness-of-fit test for each class $\mathcal{P}_i$. 
For example, if $\mathcal{P}_{i}$ is the family of Normal distributions, the unknown parameters are $\bm{\bm{\theta}} = (\mu, \sigma^2) \in\mathbb{R}^2$.

We design the statistical test by first estimating some parameters $\hat{\bm{\theta}}_i$ for each class $ \mathcal{P}_i$, so as to identify a representative \gls{cdf} $F_i(\cdot; \hat{\bm{\theta}}_i)$. Then, we perform a \gls{ks} goodness-of-fitting test for this representative distribution.
However, estimating $\hat{\bm{\theta}}_i$ from the observed data can alter the asymptotic distribution of the test statistic in Eq.~\eqref{eq:kolmogorov_statistic}, rendering it dependent on these parameters. Consequently, even under the null hypothesis, the distribution may deviate from the Kolmogorov distribution, and   require a recalibration of the critical values~\cite{babu_goodness}. 
To address this issue, we adopt a parametric Bootstrap Resampling scheme as proposed in \cite{babu_goodness, babu2006astrostatistics}, and described in the following steps.

\subsubsection{Parameter Estimation}
For each class $\mathcal{P}_i\in\mathbb{P}$, $i\in\{1,2,\dots,|\mathbb{P}|\}$, we compute the \acrfull{mle} from the observed data $B_m(\hat{\mathcal{D}})$ to estimate $\hat{\bm{\theta}}_i$. 
Then, we select CDF $F_i(\cdot;\hat{\bm{\theta}}_i)\in\mathcal{P}_i$ as a representative distribution for class $\mathcal{P}_i$, thus as the model being tested. 

\subsubsection{Target KS Statistic}
We define the composite null hypothesis of the statistical test as
\begin{equation}
    H^i_0:\text{draw samples from } F_i(\cdot;\hat{\bm{\theta}}_i), 
\end{equation}
and compute the \gls{ks} statistic as
    \begin{equation}\label{eq:ks_stat_boot}
        D(i) = \max_{x\in B_m(\hat{\mathcal{D}})}|\tilde{F}(x) - F_i(x;\hat{\bm{\theta}}_i)|,
    \end{equation}
    where $\tilde{F}$ is the empirical \gls{cdf} of the observed set $B_m(\hat{\mathcal{D}})$.  
    Notice that the hypothetical \gls{cdf} $F_i(\cdot;\hat{\bm{\theta}}_i)$ depends on $B_m(\hat{\mathcal{D}})$, and $D(i)$ is no longer distribution-free. Therefore, a Bootstrap Resampling scheme is necessary.

\begin{table}[t]
    \centering
    \caption{Parameters for the selected statistical models based on Table~\ref{tab:pvalues}.}
    \label{tab:fitting_params}
    \begin{tabular}{lllll}
    \toprule
    $m$ & Distribution $\mathbb{P}$ & Parameters & Values \\
    \midrule
    D0/S0 & tLocationScale & $\mu,\sigma,\nu$ & 3172.74, 64.41, 1.49 \\
    D0/S1 & Normal & $\mu,\sigma$ & 1458.7, 455.36 \\
    D0/S2 & Gamma & $a,b$ & 1.87, 131.97 \\
    D11/S0 & Nakagami & $\mu,\omega$ & 9.31, 4914.06 \\
    D14/S0 & Logistic & $\mu,\sigma$ & 197.54, 8.96 \\
    D14/S1 & tLocationScale & $\mu,\sigma,\nu$ & 98.11, 16.83, 4.08 \\
    D14/S2 & Gamma & $a, b$ & 2.81, 6.06 \\
    \bottomrule
\end{tabular}%
%
\end{table}

\begin{figure*}[t!]
  \centering
  \begin{subfigure}{0.49\textwidth}
  \vskip 0pt
    \centering
    \input{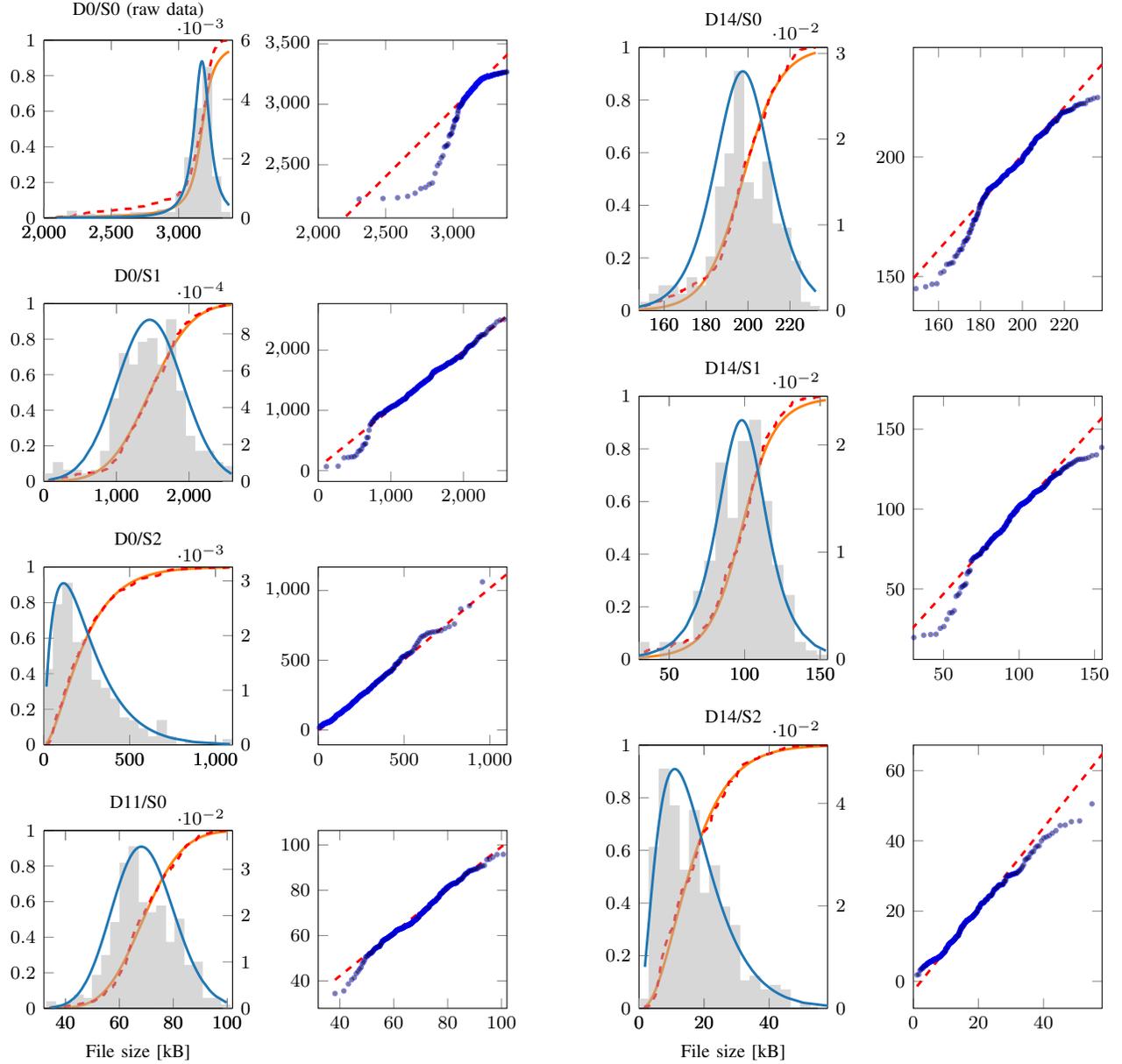}
  \end{subfigure}
  \begin{subfigure}{0.49\textwidth}
  \vskip 0pt
    \centering
   \definecolor{darkgray176}{RGB}{176,176,176}
\definecolor{darkorange}{RGB}{255,127,14}
\definecolor{steelblue}{RGB}{31,119,180}
\definecolor{whitesmoke}{RGB}{245,245,245}
\pgfplotsset{every tick label/.append style={font=\scriptsize}}
\pgfplotsset{
tick label style={font=\footnotesize},
label style={font=\footnotesize},
legend  style={font=\footnotesize}
}
\begin{tikzpicture}
\def\hratio{0.228}
\def\wratio{0.5}
\def\qqplotBisectorColor{red}



\def\binwidth{67}

\def\binwidth{4.5}
\def\xmin{148}
\def\xmax{238}
\def\filename{1450}
\begin{axis}[
    name=cdf\filename,
    at={(cdf1150.south west)},
    anchor=north west,
    yshift=-1.3cm,
    title={\footnotesize D14/S0},
    ylabel={},
    axis y line*=left,
    minor xtick={},
    xmin=\xmin,
    xmax=\xmax,
    ymin=0,
    ymax=1,
    width=\wratio\textwidth,height=\hratio\textheight,
]
\addplot[mark=no, color=darkorange, line width=1.1pt] table [x=x, y=y, col sep=comma]{Figures/Fitting_Plots/data/\filename_cdf.csv};
\end{axis}

\begin{axis}[
    at={(cdf\filename.north west)},
    anchor=north west,
    ylabel={},
    axis y line=none,
    axis x line=none,
    minor xtick={},
    xmin=\xmin,
    xmax=\xmax,
    ymin=0,
    ymax=1,
    width=\wratio\textwidth,height=\hratio\textheight,
]
\addplot[mark=no, color=red, line width=1.1pt, dashed] table [x=x, y=y, col sep=comma]{Figures/Fitting_Plots/data/\filename_ecdf.csv};
\end{axis}

\begin{axis}[
    at={(cdf\filename.north west)},
    anchor=north west,
    area style,
    minor xtick={},
    axis y line=none,
    xmin=\xmin,
    xmax=\xmax,
    ymin=0,
    width=\wratio\textwidth,height=\hratio\textheight,
]
\addplot+[ybar, mark=no, color=darkgray176, opacity=0.5, bar width=\binwidth] plot table [x=x, y=y, col sep=comma]{Figures/Fitting_Plots/data/\filename_hist.csv};
\end{axis}

\begin{axis}[
    at={(cdf\filename.north west)},
    anchor=north west,
    axis y line*=right,
    axis x line=none,
    ylabel={PDF},
    ylabel style ={darkorange},
    minor xtick={},
    ylabel={},
    xmin=\xmin,
    xmax=\xmax,
    ymin=0,
    width=\wratio\textwidth,height=\hratio\textheight,
]
\addplot[mark=no, color=steelblue, , line width=1.1pt] table [x=x, y=y, col sep=comma]{Figures/Fitting_Plots/data/\filename_pdf.csv};
\end{axis}

\begin{axis}[
    at={(cdf\filename.north east)},
    anchor=north west,
    xshift=1.3cm,
    minor xtick={},
    ylabel={},
    xmin=\xmin,
    xmax=\xmax,
    width=\wratio\textwidth,height=\hratio\textheight,
]
\addplot[only marks, mark size=1pt, opacity=0.5, draw=blue, ] table [x=x, y=y, col sep=comma]{Figures/Fitting_Plots/data/\filename_qqplot.csv};
\addplot[dashed, mark=no, color=\qqplotBisectorColor, line width=1.1pt] table [x=x, y=y_ref, col sep=comma]{Figures/Fitting_Plots/data/\filename_qqplot.csv};
\end{axis}

\def\binwidth{7.4}
\def\xmin{0}
\def\xmin{30}
\def\xmax{155}
\def\filename{1451}
\begin{axis}[
    name=cdf\filename,
    at={(cdf1450.south west)},
    anchor=north west,
    yshift=-1.3cm,
    title={\footnotesize D14/S1},
    ylabel={},
    axis y line*=left,
    minor xtick={},
    xmin=\xmin,
    xmax=\xmax,
    ymin=0,
    ymax=1,
    width=\wratio\textwidth,height=\hratio\textheight,
]
\addplot[mark=no, color=darkorange, line width=1.1pt] table [x=x, y=y, col sep=comma]{Figures/Fitting_Plots/data/\filename_cdf.csv};
\end{axis}

\begin{axis}[
    at={(cdf\filename.north west)},
    anchor=north west,
    ylabel={},
    axis y line=none,
    axis x line=none,
    minor xtick={},
    xmin=\xmin,
    xmax=\xmax,
    ymin=0,
    ymax=1,
    width=\wratio\textwidth,height=\hratio\textheight,
]
\addplot[mark=no, color=red, line width=1.1pt, dashed] table [x=x, y=y, col sep=comma]{Figures/Fitting_Plots/data/\filename_ecdf.csv};
\end{axis}

\begin{axis}[
    at={(cdf\filename.north west)},
    anchor=north west,
    area style,
    minor xtick={},
    axis y line=none,
    xmin=\xmin,
    xmax=\xmax,
    ymin=0,
    width=\wratio\textwidth,height=\hratio\textheight,
]
\addplot+[ybar, mark=no, color=darkgray176, opacity=0.5, bar width=\binwidth] plot table [x=x, y=y, col sep=comma]{Figures/Fitting_Plots/data/\filename_hist.csv};
\end{axis}

\begin{axis}[
    at={(cdf\filename.north west)},
    anchor=north west,
    axis y line*=right,
    axis x line=none,
    ylabel={PDF},
    ylabel style ={darkorange},
    minor xtick={},
    ylabel={},
    xmin=\xmin,
    xmax=\xmax,
    ymin=0,
    width=\wratio\textwidth,height=\hratio\textheight,
]
\addplot[mark=no, color=steelblue, , line width=1.1pt] table [x=x, y=y, col sep=comma]{Figures/Fitting_Plots/data/\filename_pdf.csv};
\end{axis}

\begin{axis}[
    at={(cdf\filename.north east)},
    anchor=north west,
    xshift=1.3cm,
    minor xtick={},
    ylabel={},
    xmin=\xmin,
    xmax=\xmax,
    width=\wratio\textwidth,height=\hratio\textheight,
]
\addplot[only marks, mark size=1pt, opacity=0.5, draw=blue, ] table [x=x, y=y, col sep=comma]{Figures/Fitting_Plots/data/\filename_qqplot.csv};
\addplot[dashed, mark=no, color=\qqplotBisectorColor, line width=1.1pt] table [x=x, y=y_ref, col sep=comma]{Figures/Fitting_Plots/data/\filename_qqplot.csv};
\end{axis}

\def\binwidth{3.1}
\def\xmin{0}
\def\xmax{58}
\def\filename{1452}
\begin{axis}[
    name=cdf\filename,
    at={(cdf1451.south west)},
    anchor=north west,
    yshift=-1.3cm,
    title={\footnotesize D14/S2},
    ylabel={},
    axis y line*=left,
    minor xtick={},
    xmin=\xmin,
    xmax=\xmax,
    ymin=0,
    ymax=1,
    width=\wratio\textwidth,height=\hratio\textheight,
]
\addplot[mark=no, color=darkorange, line width=1.1pt] table [x=x, y=y, col sep=comma]{Figures/Fitting_Plots/data/\filename_cdf.csv};
\end{axis}

\begin{axis}[
    at={(cdf\filename.north west)},
    anchor=north west,
    ylabel={},
    axis y line=none,
    axis x line=none,
    minor xtick={},
    xmin=\xmin,
    xmax=\xmax,
    ymin=0,
    ymax=1,
    width=\wratio\textwidth,height=\hratio\textheight,
]
\addplot[mark=no, color=red, line width=1.1pt, dashed] table [x=x, y=y, col sep=comma]{Figures/Fitting_Plots/data/\filename_ecdf.csv};
\end{axis}

\begin{axis}[
    at={(cdf\filename.north west)},
    xlabel={File size [kB]},
    anchor=north west,
    area style,
    minor xtick={},
    axis y line=none,
    xmin=\xmin,
    xmax=\xmax,
    ymin=0,
    width=\wratio\textwidth,height=\hratio\textheight,
]
\addplot+[ybar, mark=no, color=darkgray176, opacity=0.5, bar width=\binwidth] plot table [x=x, y=y, col sep=comma]{Figures/Fitting_Plots/data/\filename_hist.csv};
\end{axis}

\begin{axis}[
    at={(cdf\filename.north west)},
    anchor=north west,
    axis y line*=right,
    axis x line=none,
    ylabel={PDF},
    ylabel style ={darkorange},
    minor xtick={},
    ylabel={},
    xmin=\xmin,
    xmax=\xmax,
    ymin=0,
    width=\wratio\textwidth,height=\hratio\textheight,
]
\addplot[mark=no, color=steelblue, , line width=1.1pt] table [x=x, y=y, col sep=comma]{Figures/Fitting_Plots/data/\filename_pdf.csv};
\end{axis}

\begin{axis}[
    at={(cdf\filename.north east)},
    anchor=north west,
    xshift=1.3cm,
    minor xtick={},
    ylabel={},
    xmin=\xmin,
    xmax=\xmax,
    width=\wratio\textwidth,height=\hratio\textheight,
]
\addplot[only marks, mark size=1pt, opacity=0.5, draw=blue, ] table [x=x, y=y, col sep=comma]{Figures/Fitting_Plots/data/\filename_qqplot.csv};
\addplot[dashed, mark=no, color=\qqplotBisectorColor, line width=1.1pt] table [x=x, y=y_ref, col sep=comma]{Figures/Fitting_Plots/data/\filename_qqplot.csv};
\end{axis}

\end{tikzpicture}
  \end{subfigure}
      \caption{Fitting plots: data histogram (light grey), empirical \gls{cdf} (dashed red), fitted \acrshort{pdf} (light blue), fitted \gls{cdf} (orange), and QQ-plots (right).}
    \label{fig:fitting_plots}
\end{figure*}

\subsubsection{Parametric Bootstrap Resampling scheme}\label{sec:boot_resampling_scheme}
We define $L$ independent Bootstrap resamples $\{ B^*_{m,l}(\hat{\mathcal{D}}) \}_{l=1}^L$ from the estimated population $F_i(\cdot;\hat{\bm{\theta}}_i)$ and, for each $B^*_{m,l}(\hat{\mathcal{D}})$, we compute the \gls{mle} to estimate $\hat{\bm{\theta}}^*_{i,l}$. 
Our goal is to obtain $L$ \gls{ks} statistics, and compare them with the target statistics in Eq.~\eqref{eq:ks_stat_boot}. 
We denote as $\tilde{F}_{{l}}(x)$ the empirical \gls{cdf} of $B^*_{m,l}(\hat{\mathcal{D}})$. 
Then, the \gls{ks} statistic computed on the $l$-th Bootstrap resample~is 
\begin{align}
    D_l(i) = \max_{x\in B^*_{m,l}(\hat{\mathcal{D}})}&|\tilde{F}_{l}(x) - F_i(x;\hat{\bm{\theta}}^*_{i,l})|.
\end{align}
 Since $\sqrt{N}\big(\tilde{F}(x)- F_i(x;\hat{\bm{\theta}}_i)\big)$ and  $\sqrt{N}\big(\tilde{F}_{l}(x) - F_i(x;\hat{\bm{\theta}}^*_{i,l})\big)$ converge to the same Gaussian process, $\sqrt{N}D(i)$ and $\sqrt{N}D_{l}(i)$ have the same limiting distribution \cite{babu2006astrostatistics}. 
 Therefore, the critical values relative to Eq.~\eqref{eq:ks_stat_boot} can be obtained computing the ($1-\alpha$)-percentile of $\big\{ \sqrt{N}D_{l}(i) \big\}_{l=1}^L$, 
 where $\alpha$ is the significance level.
 Equivalently, we can compute the Bootstrap p-value $\pi_{i,m}^*$ as 
\begin{align}\label{eq:pvalues}
    \pi_{i,m}^* &= \frac{1}{L}\sum_{l=1}^L \mathds{1}_{\{D_{l}(i)>D(i)\}}\\
    &=1-\tilde{F}_{D^*(i)}(D(i))\approx \text{Prob}\big(D^{*}(i)>D(i)\big),
\end{align}
where $\tilde{F}_{D^*(i)}(D(i))$ is the empirical \gls{cdf} of $\big\{ D_{l}(i) \big\}_{l=1}^L$. Therefore, $H^i_0$ can be rejected at significance level $\alpha$ if $\pi_{i,m}^*<\alpha$, i.e., $\text{Prob}\big(D^*(i)\leq D(i)\big)>1-\alpha$.

\subsubsection{Model Selection}
The \gls{ks} statistical test with the Bootstrap Resampling scheme described above provides, for each configuration $m$, $|\mathbb{P}|$ statistical test outcomes, one for each distribution family. We define the set of indices of distribution families for which the null hypothesis cannot be rejected as $\mathcal{J}_m$. 
Then, we select as best fit the CDF family $\mathcal{P}_{i_m^{\star}}$ and $F_{i^{\star}_m}(\cdot; \hat{\bm{\theta}}_{i_m^\star})$, $\forall \, i\in\mathcal{J}_m$, such that $i_m^\star = \arg\max_{i\in\mathcal{J}_m}\pi_{i,m}^*$. If $\mathcal{J}_m=\emptyset$, we select as best fit the model that minimizes the \gls{nrmse}, computed as 
\begin{equation}
    \text{NRMSE}=\sqrt{\frac{ \sum_{k=1}^N\big( \tilde{F}(x_k) - F_i(x_k;\hat{\bm{\theta}}_i)\big)^2 }{ \sum_{k=1}^N\big( \tilde{F}(x_k) - \bar{F_i}(\hat{\bm{\theta}}_i) \big)^2 }}, %
\end{equation}
where $\bar{F_i}(\hat{\bm{\theta}}_i) = \frac{1}{N}\sum_{k=1}^N F_i(x_k;\hat{\bm{\theta}}_i)$.

\subsection{Statistical Results}
\label{ssub:statistical-results}
In this section we present and discuss the results obtained from the statistical method in Sec.~\ref{ssub:statistical_test}. 
Table~\ref{tab:pvalues} shows the Bootstrap p-values $\pi^*_{i,m}$ computed according to Eq.~\eqref{eq:pvalues} for each compression configuration $m$, and considering several \gls{cdf} families. We set $\alpha=0.01$ and $L=1000$. 
We represent in bold the p-values that pass the test, i.e., $\pi_{i,m}^*\in\mathcal{I}_m$, and highlight in grey the entries selected as best fits, that is $i_m^\star$. 
We see that the size of LiDAR point clouds can be represented as tLocationScale, Nakagami, Normal and Logicstic distributions, depending on the compression configuration.
Notice that D0/S0 is the only model that does not pass the test ($\mathcal{I}_m=\emptyset$). 
In this case,  we choose as best fit the distribution with the minimum \gls{nrmse}, i.e., tLocationScale, for which \gls{nrmse} $=7.4\cdot 10^{-3}$. 
Finally, in Table~\ref{tab:fitting_params} we report the parameters $\hat{\bm{\theta}}_{i^\star}$ for the best fitting models, that will be used in the ns-3 implementation (Sec.~\ref{sec:ns3}) and the end-to-end performance evaluation (Sec.~\ref{sec:performance_evaluation}).

Moreover, we complement this numerical analysis with fitting plots. 
Specifically, for each model, Fig.~\ref{fig:fitting_plots} illustrates the empirical \gls{cdf} $\tilde{F}$ and \gls{pdf} of the observed data, as well as the validated theoretical \gls{cdf} $F_{i^\star}(\cdot;\hat{\bm{\theta}}_{i^\star})$, together with the QQ-plot (Quantile-Quantile plot). The latter serves as a graphical diagnostic method to visually compare two distributions. It plots on the x-axis the quantile function of one distribution and on the y-axis the quantile function of the other distribution. Therefore, if two distributions are identical, the QQ-plot will lie along the bisector of the first quadrant.
We see that the \gls{pdf} and \gls{cdf} of all the models generally fit the empirical data, confirming the accuracy of our statistical results.
The QQ-plots also show good accuracy, with only minor deviations in the tails, which decrease as the semantic level (S0, S1, S2) increases. 
This is probably due to the fact that, as fewer points remain in the point cloud, the structure of the resulting data is less complex, which facilitates a better fit to a statistical distribution.
The only exception is for the D0/S0 model, where the QQ-plot reveals a heavy left tail, which highlights some discrepancy between the statistical and empirical distributions. 
This is consistent with the fact that D0/S0 is the only model that fails the~test.

\section{ns-3 Implementation of Statistical Models}
\label{sec:ns3}
In this paper, we measure the accuracy of the proposed statistical models for the size of automotive data based on their impact on network metrics.
To do so, we use ns-3 as a default system-level simulator~\cite{henderson2008network}.
Notably, ns-3 has gained great popularity within the network simulation community. It consists of a large set of predefined, scalable, ready-to-use, open-access modules to simulate different parts of the (wireless) network. It also comes with modules to simulate \gls{v2x} networks based on the most recent \gls{3gpp} specification for NR V2X~\cite{drago2020millicar}, mobility traces using SUMO~\cite{krajzewicz2012recent}, and a pipeline to simulate and test machine learning algorithms within the \gls{ran}~\cite{drago2022artificial}. As such, it stands out as one of the most complete 5G-oriented tools to perform accurate simulations in the context of vehicular networks.
In Sec.~\ref{sub:implementation_of_random_distributions} we describe the ns-3 application model, and in Sec.~\ref{sub:ns3math} we focus on the implementation of the statistical models in ns-3, which we validate in Sec.~\ref{sub:implementation_evaluation}.

\subsection{Application Model} 
\label{sub:implementation_of_random_distributions}
The statistical models have been implemented at the application layer of ns-3 in the new \emph{StatisticalTraffic} module, which we made publicly available.\footnote{The source code of the ns-3 \emph{StatisticalTraffic} module: \url{https://github.com/signetlabdei/kitti-statistical-dataset}.}
Given the nature of point clouds, it is built on top of the \emph{BurstyApplication}~\cite{bursty}, originally designed to model \gls{xr} traffic.
Specifically, it serves as a bursty traffic generator that produces bursts of network packets based on two input parameters: the size of LiDAR point clouds, modeled based on the statistical distributions presented in Sec.~\ref{ssub:statistical-results}, and a predefined inter-arrival time.

Notably, the proposed \emph{StatisticalTraffic} module simulates the packet generation process for the seven different \gls{hsc} compression configurations presented in Sec.~\ref{sec:compression_model_refinement}. 
First, the application selects the point cloud size distribution through the \texttt{GetModel} routine. Then, the \texttt{StatisticModelInit} routine initializes the distribution parameters according to Table~\ref{tab:fitting_params}.
Finally, the \texttt{BurstSize} and \texttt{FramePeriod} are generated based on the corresponding statistical models, and passed to the \textit{BurstyApplication} through the \texttt{BurstGenerator} interface.
Fig.~\ref{fig:overview} depicts a block diagram of the ns-3 module. 

The remaining components of the \gls{5g} NR V2X protocol stack are emulated using the \emph{mmWave} module~\cite{mezzavilla2018end}.
This module features a customized \gls{phy} layer that accommodates 5G \gls{nr} frame formats and numerologies, as well as a \gls{mac} layer supporting ad hoc beamforming and scheduling strategies.
The \gls{pdcp} layer leverages the ns-3 \emph{Lena} module for \gls{lte} networks~\cite{piro2011lte}, providing network functions such as packet segmentation, retransmissions, and reassembly.
Furthermore, this module facilitates non-standalone deployments, handovers, and mobility management through dual connections, along with \gls{ca} at the \gls{mac} layer.

\subsection{Implementation Details of Statistical Distributions}
\label{sub:ns3math}
The implementation of \emph{StatisticalTraffic} requires the integration of statistical distributions directly within the ns-3 framework.
Although ns-3 offers built-in support for a wide range of random variables, it does not implement the tLocationScale, Logistic and Nakagami distributions, which are essential for the proposed traffic models.
For tLocationScale and Logistic, we use the \gls{icdf} sampling method, a well-known statistical technique used to generate random samples from a probability distribution with a known \gls{cdf}. It involves mapping uniformly distributed random numbers (typically between 0 and 1) to specific quantiles of the desired distribution using the inverse of the \gls{cdf}. 
For Nakagami, we exploit the fact that it can be expressed as a function of a Gamma distribution. 
Below, we provide a mathematical description for these distributions.

\subsubsection{tLocationScale}
    The direct implementation of the \gls{icdf} for tLocationScale is known to be NP-hard. 
    Consequently, various alternative approaches have been proposed~\cite{shaw2006sampling}. In our implementation, we utilize the central power series method. Thus, we represent the \gls{icdf} of tLocationScale as 
    \begin{equation}\label{eq:tloc}
         F^{-1}_{\textrm{tloc}}(u;\mu,\sigma,\nu) = \Bigg(\nu+\sum_{i=1}^{\infty} c(i,\nu)V(\nu,u)^{2i +1}\Bigg)\sigma + \mu,
    \end{equation}
    where $\mu$, $\sigma$ and $\nu$ are the location, scale and shape parameters in Table~\ref{tab:fitting_params}, and $V(\nu,u)$ is determined according to 
    \begin{equation}\label{eq:tloc_v}
    V(\nu,u) = \sqrt{\nu\pi}\Big(u - \frac{1}{2}\Big) \frac{\Gamma(\nu/2)}{\Gamma((\nu+1)/2)},
    \end{equation}
    where $c(i,\nu)$ represents the coefficients of the power series reported in \cite[Sec. 5]{shaw2006sampling}.

\subsubsection{Logistic}
    In order to implement \gls{icdf} of a {Logistic} distribution, we follow the process outlined in~\cite{le2010performance}. 
    Specifically, the \gls{icdf} can be expressed as a function of the location and scale parameters $\mu$ and $\sigma$ in Table~\ref{tab:fitting_params} as 
        \begin{equation}\label{inversecdflogistic}
        F^{-1}_{\textrm{log}}(u; \mu,\sigma)= \mu-\sigma\ln{\frac{1-u}{u}}. 
    \end{equation}

    Thus, a sample from a Logistic distribution can be generated by drawing a number according to a uniform distribution $u\sim\mathcal{U}(0,1)$, and mapping it through Eq.~\eqref{inversecdflogistic}.

\subsubsection{Nakagami}
    The CDF of a Nakagami distribution can be explicitly derived from the Gamma distribution $F_{\Gamma}$, which is natively implemented in ns-3. 
    Therefore, we use a random number generator for $F_{\Gamma}$, and adapt it according to
    \begin{equation}
    \label{eq:nakagami}
        F_{\textrm{nak}}(x;\mu,\omega)= \sqrt{F_{\Gamma}\Big(x;\mu,\frac{\omega}{\mu}\Big)},
    \end{equation}
    where $\mu$ and $\omega$ represent the shape and spread parameters in Table~\ref{tab:fitting_params}, respectively.

\begin{table}[t]
\caption{P-values of the implemented distributions in ns-3. We represent in bold the p-values that pass the test.}
\label{Tab:ns3-evalution}
\centering
\scriptsize
\begin{tabular}{ll}
    \toprule
    $\mathbb{P}$ & p-value \\ \midrule
    tLocationScale ($\sim$ \gls{icdf} sampling method) & \textbf{0.296} \\
    Logistic ($\sim$ \gls{icdf} sampling method)   & \textbf{0.880} \\
    Nakagami ($\sim$ Gamma CDF) & \textbf{0.702} \\
\bottomrule
\end{tabular}
\end{table}

\begin{table}[t]
\caption{Network simulation parameters.}
\label{Tab:parameters}
\centering
\scriptsize
\begin{tabular}{ll}
    \toprule
    Parameter & Value \\ \midrule
    Carrier frequency & 28~GHz \\
    Bandwidth    & 200~MHz \\
    Transmit Power & 30 dBm \\
    Channel Model & \makecell[l]{3GPP TR 38.901 (UMi-Street Canyon) \cite{3gpp.38.901}}\\
    \gls{lidar} Inter Burst Interval & 100~ms \\
    Buffer Size & 12 MB\\
    Data direction & Uplink\\
\bottomrule
\end{tabular}
\end{table}

\begin{table}[t]
    \centering
    \caption{Mean PRR vs. $d$.}
    \label{tab:prr}
    \resizebox{\columnwidth}{!}{%
    \begin{tabular}{lcccccccccc}
        \toprule
         $d$ [m] & 15-45  & 75  & 105 & 135  & 165  & 195  & 225  & 255   & 300 \\\midrule
         PRR & 100.0 & 82.8 & 74.5  & 65.4  & 48.3  & 41.2  & 34.3  & 16.8 & 14.0\\
         \bottomrule
    \end{tabular}
    }
\end{table}

\begin{figure}[t]
    \setlength\fheight{.5\columnwidth}
    \setlength\fwidth{\columnwidth}
    \centering
    \input{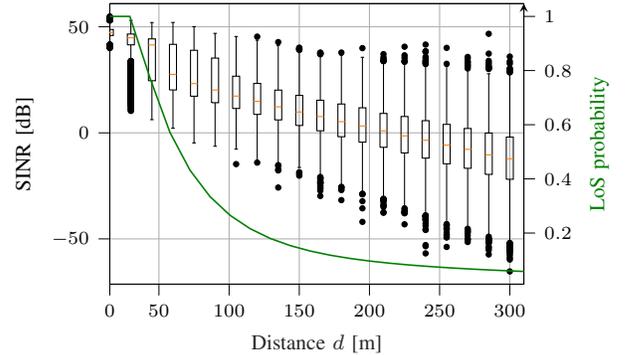}
    \caption{\acrshort{sinr} and \acrshort{los} probability (based on the model in~\cite{3gpp.38.901}) vs. $d$.}
    \label{fig:sinr}
\end{figure}

\subsection{Implementation Validation}
\label{sub:implementation_evaluation}
We test our implemented ns-3 distributions against the corresponding built-in distributions in the Python module \emph{Scipy}. Specifically, we perform a \gls{ks} test at a significance level $\alpha=0.05$ to assess whether a sample drawn from our ns-3 implementation of the tLocationScale, Logistic, and Nakagami distributions matches the one from \emph{Scipy}. 
The p-values of the tests are reported in Table~\ref{Tab:ns3-evalution}. We can see that, in all the models, the null hypothesis cannot be rejected since the p-values are far greater than the significance level, which validates the accuracy of our ns-3 implementations.

\section{Network Performance Evaluation of Statistical Models}\label{sec:performance_evaluation}
While in Sec.~\ref{ssub:statistical-results} we validated the proposed traffic models via statistical tests, we now evaluate their accuracy based on the effect on some network metrics such as latency and throughput via ns-3 simulations.
The goal is to verify that statistical models can effectively represent an alternative to real data, provided that the network performance remains consistent in terms of both values and overall trends.
 In Sec.~\ref{sub:simulation_setup} we describe our simulation setup and parameters, while in Sec.~\ref{sub:numerical_results} we present our numerical results.

\subsection{Simulation Setup}
\label{sub:simulation_setup}

We consider a simple yet realistic urban scenario, where a vehicle transmits \gls{lidar} data to a \gls{gnb} through a \gls{mmwave} link at  28 GHz.
The two nodes have the same height, and the distance $d$ between them is increased from 15 to 300~m as the vehicle moves away from the \gls{gnb}.

We run ns-3 simulations using the parameters reported in \cref{Tab:parameters}.
The channel is simulated according to the \gls{3gpp} TR 38.901 UMi-Street Canyon model~\cite{3gpp.38.901}.
Both the \gls{gnb} and the vehicle are equipped with an $8\times 8$ \gls{upa}, and 
the beamforming vectors are computed based on the \gls{svd} of the channel matrix.

At the application layer, \gls{lidar} bursts are generated every 100~ms.
The data is compressed using \gls{hsc} at the \gls{pdcp} layer according to the 7 representative configurations introduced in~\cref{ssub:final_selection}: D0/S0, D0/S1, D0/S2, D11/S0, D14/S0, D14/S1, D14/S2.
We measure the following \gls{e2e} metrics at the \gls{pdcp} layer: (i)
    {\gls{e2e} throughput}, measured as the ratio between the number of bytes received over the entire simulation and the total simulation time; and (ii)
     {\gls{e2e} latency}, measured as the difference between the time at which each packet is generated at the application layer and when it is successfully received
    (which accounts for the transmission, compression, queuing, and decompression times).

\subsection{Simulation Results}
\label{sub:numerical_results}

\begin{table*}[t]
    \caption{(Measured) encoding/decoding/inference time, and (simulated) source rate and throughput for different \gls{hsc} compression configurations.}
    \label{tab:avg_thr}
    \centering
    \scriptsize
    \begin{tabular}{lccccccc}
        \toprule
        {Parameter} & D0/S0 & D0/S1 & D0/S2 & D11/S0 & D14/S0 & D14/S1 & D14/S2  \\ \midrule
        Avg. size [MB] & 3.204 & 1.511 & 0.235  & 0.071 & 0.202 & 0.100 & 0.016 \\
        Encoding time [ms] & 0 & 0 & 0 & 23.3  & 28.2 & 12.97 & 1.95 \\
        Decoding time [ms] & 0  & 0  & 0 & 10.48  & 13.57  & 5.81 & 0.72\\
        RangeNet++ inference time [ms] & 0  & 56  & 56 & 0  & 0  & 56 & 56\\ \midrule
        Source rate  [Mbps] & 256.3 & 120.9 & 18.8 & 5.7 & 16.2 & 8 & 1.3\\
        Throughput ($d<50$ m)  [Mbps] & 259.4 & 124.3 & 20.2 & 5.6 & 17.3 & 8.6 & 1.4\\
        Throughput ($d>50$ m) [Mbps]  & 151 & 72.2 & 11.3 & 2.9 & 10 & 5 & 0.7\\
        \bottomrule
    \end{tabular}
\end{table*}

\begin{figure*}[t]
    \setlength\fheight{.6\columnwidth}
    \setlength\fwidth{2\columnwidth}
    \centering
    \input{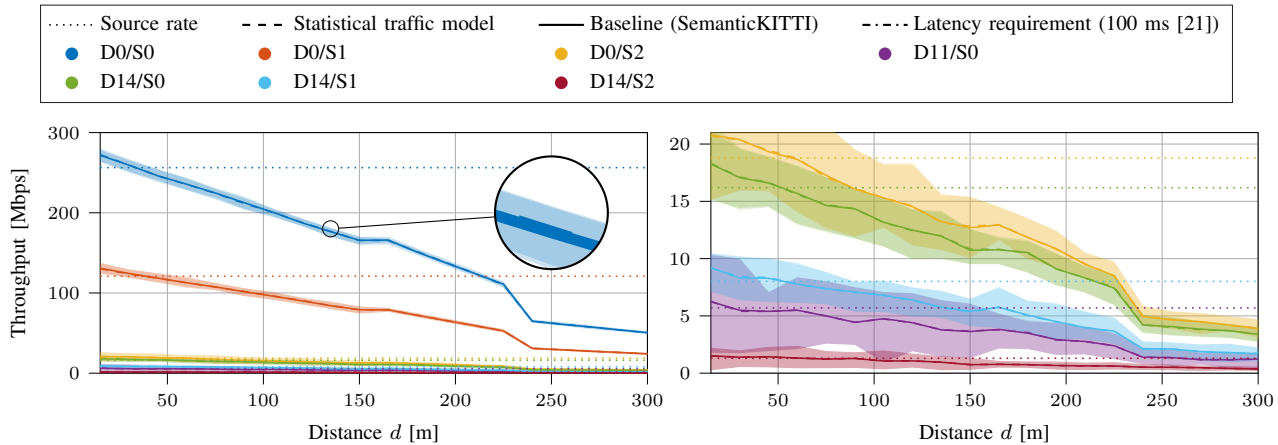}
    \caption{Mean throughput and confidence intervals (shaded areas) vs. $d$, and for different \gls{hsc} compression configurations.
Solid (dashed) lines are relative to the use of real data from SemanticKITTI (statistical traffic models).}
    \label{fig:throughput}
\end{figure*}

In this section, we compare the network performance obtained using real data from the SemanticKITTI dataset~\cite{behley2019semantickitti}, vs. considering our statistical traffic models.
Hence, the data size of the point clouds, whether extracted directly from SemanticKITTI or sampled from random distributions, is used in the \textit{BurstyApplication} or \textit{StatisticalTraffic} modules in ns-3, respectively, which in turn generates a stream of packets according to the procedure described in \cref{sec:ns3}.

\subsubsection{SINR and PRR} 
\label{ssub:sinr_and_prr}


In \cref{fig:sinr} we plot the average \gls{sinr} as a function of $d$.
We observe that, as expected, at short distances the \gls{sinr} is high (around 50~dB), as the channel is mainly in \gls{los} conditions.
As the vehicle moves away from the \gls{gnb}, the \gls{los} probability (green line) decreases according to the model in~\cite{3gpp.38.901}, and the SINR drops down to $-12$~dB at 300~m.
This translates into a slow degradation of the theoretical capacity of the channel that, considering the average ($25\%$ quantile) \gls{sinr}, varies between 3.14 (3)~Gbps, 2.75 (1.63)~Gbps, and 16.42 (1.8)~Mbps at 15, 45, and 300~m, respectively.
These results are also confirmed by the \gls{prr} reported in \cref{tab:prr}, which decreases as $d$ increases.
For $d<45$~m, all the packets are successfully delivered; however, as $d>45$~m, the packet loss increases due to the worse channel conditions.

\begin{figure}
    \setlength\fheight{0.5\columnwidth}
    \setlength\fwidth{\columnwidth}
    \centering
    \input{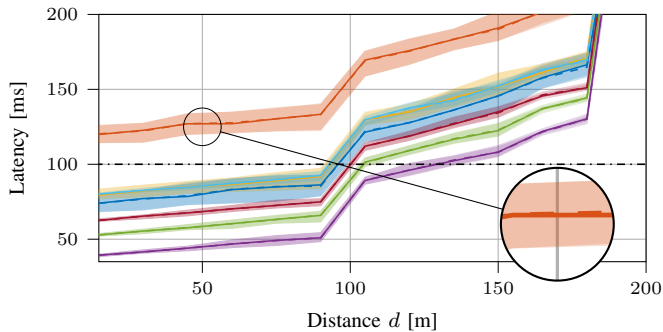}
    \caption{Mean latency and confidence intervals (shaded areas) vs. $d$, and for different \gls{hsc} compression configurations.
Solid (dashed) lines are relative to the use of real data from SemanticKITTI (statistical traffic models).}
    \label{fig:latency}
\end{figure}

\subsubsection{Throughput and latency} 

\cref{tab:avg_thr} reports the average source rate of the application, calculated by multiplying the average data size by the rate of the \gls{lidar} (10~Hz).\footnote{Note that, unlike the throughput, the source rate does not include the headers nor the communication overhead introduced by the protocol stack. This explains why the throughput can be slightly larger than the source rate in some cases (e.g., D0/S2, D14/S0$-$2).}
The results indicate that existing V2X technologies cannot support the source rate of raw data (256.3 Mbps for D0/S0) or when conservative compression is applied (e.g., 120.9 Mbps for D0/S1). 
For comparison, the peak nominal throughput of the  IEEE 802.1p protocol is only  27~Mbps~\cite{interworking2016abboud}, which motivates the need for compression and/or more advanced V2X solutions like NR V2X operating at \glspl{mmwave}~\cite{zugno2020toward}.

In \cref{fig:throughput,fig:latency} we plot the \gls{e2e} throughput and latency, respectively, vs. $d$.
The dashed lines are for the results obtained using real data from SemanticKITTI, while solid lines correspond to using statistical traffic models. 
We clearly see that there is an almost perfect overlap between the two sets of curves for all \gls{hsc} compression configurations and values of $d$.
We conclude that replacing real data with statistical traffic models has a negligible impact on the network.
Interestingly, this is true also for model D0/S0, that did not pass the KS test in Sec.~\ref{ssub:statistical-results}; although this model is not formally accurate to characterize the size of LiDAR data, it remains sufficiently precise in terms of network metrics.

Moreover, we see that the throughput (latency) decreases (increases) as $d$ increases, which is due to the lower SINR at long distance.
From~\cref{fig:latency}, we see that the latency is always below the application requirement for autonomous driving (set to 100~ms based on 3GPP specifications~\cite{3gpp.22.186}) for all compression configurations except D0/S1 and when $d>100$~m.
This may seem counterintuitive, considering that D0/S1 applies more compression than D0/S0 (raw data), so data transmission should be faster in principle.
However, for D0/S1, the additional inference time of RangeNet++ (i.e., 56 ms as reported in \cref{tab:avg_thr}), combined with the encoding and decoding times for compression, must also be considered, which is not the case for D0/S0.
The performance further improves for D11/S0, suggesting that reducing the number of quantization bits is desirable for both throughput and latency.
Conversely, increasing the SL from S0 to S1 and S2 improves the throughput, at the cost of latency, again due to the additional inference time required for segmentation. 
Nevertheless, at short distance, the latency remains under the 100~ms threshold.

Finally, notice that a more aggressive compression configuration might inevitably degrade the quality of the point cloud (generally measured in terms of the mean Average Precision (mAP)), and affect the performance of object detection algorithms at the receiver.
This analysis was partially addressed in \cite{mason2022reinforcement,bragato2024federated}, and is out of the scope of this paper.

\begin{table}[t]
    \centering
    \resizebox{\columnwidth}{!}{%
    \begin{tabular}{cccccccc}
    \toprule
    {$d$~[m]}     &     {D0/S0}  & {D0/S1}  & {D0/S2}  & {D11/S0}  & {D14/S0}  & {D14/S1}  & {D14/S2}\\
         \midrule
 15 & \cmark &   \cmark &   \cmark &   \cmark &   \cmark &   \cmark &   \cmark \\   
 30 & \cmark &   \cmark &   \cmark &   \cmark &   \cmark &   \cmark &   \cmark \\   
 45 & \cmark &   \cmark &   \cmark &   \cmark &   \cmark &   \cmark &   \cmark \\   
 60 & \cmark &   \cmark &   \cmark &   \cmark &   \cmark &   \cmark &   \cmark \\   
 75 & \cmark &   \cmark &   \cmark &   \cmark &   \cmark &   \cmark &   \cmark \\   
 90 & \cmark &   \cmark &   \cmark &   \cmark &   \cmark &   \cmark &   \cmark \\   
105 & \cmark &   \cmark &   \cmark &   \cmark &   \cmark &   \cmark &   \cmark \\   
120 & \cmark &   \cmark &   \cmark &   \cmark &   \cmark &   \cmark &   \cmark \\   
135 & \cmark &   \cmark &   \cmark &   \cmark &   \cmark &   \cmark &   \cmark \\   
165 & \xmark &  \cmark &   \cmark &   \cmark &   \cmark &   \cmark &   \cmark \\   
195 & \xmark &  \cmark &   \cmark &   \cmark &   \cmark &   \cmark &   \cmark \\   
225 & \xmark &  \cmark &   \cmark &   \cmark &   \cmark &   \cmark &   \cmark \\   
255 & \xmark &  \cmark &   \cmark &   \cmark &   \cmark &   \cmark &   \cmark \\   
285 & \cmark &   \cmark &   \cmark &   \cmark &   \cmark &   \cmark &   \cmark \\   
300 & \xmark &  \cmark &   \cmark &   \cmark &   \cmark &   \cmark &   \cmark \\   
\bottomrule
    \end{tabular}%
    }
    \caption{Results of the KS test for the \gls{e2e} throughput obtained using the statistical models and real data with the SemanticKITTI dataset. The mark \cmark (\xmark) is used if the test is passed (not passed).}
    \label{tab:ks_thr}
\end{table}

\begin{table}[t]
    \centering
    \resizebox{\columnwidth}{!}{%
    \begin{tabular}{cccccccc}
    \toprule
    {$d$~[m]}     &     {D0/S0}  & {D0/S1}  & {D0/S2}  & {D11/S0}  & {D14/S0}  & {D14/S1}  & {D14/S2}\\\midrule
 15 & \cmark &   \cmark &   \cmark &   \cmark &   \cmark &   \cmark &   \cmark \\   
 30 & \cmark &   \cmark &   \cmark &   \cmark &   \cmark &   \cmark &   \cmark \\   
 45 & \cmark &   \cmark &   \cmark &   \cmark &   \cmark &   \cmark &   \cmark \\   
 60 & \cmark &   \cmark &   \cmark &   \cmark &   \cmark &   \cmark &   \cmark \\   
 75 & \cmark &   \cmark &   \cmark &   \cmark &   \cmark &   \cmark &   \cmark \\   
 90 & \cmark &   \cmark &   \cmark &   \cmark &   \cmark &   \cmark &   \cmark \\   
105 & \cmark &   \cmark &   \cmark &   \cmark &   \cmark &   \cmark &   \cmark \\   
120 & \cmark &   \cmark &   \cmark &   \cmark &   \cmark &   \cmark &   \cmark \\   
135 & \cmark &   \cmark &   \cmark &   \cmark &   \cmark &   \cmark &   \cmark \\   
165 & \cmark &   \cmark &   \cmark &   \xmark &  \cmark &   \xmark &  \xmark \\
195 & \cmark &   \cmark &   \cmark &   \cmark &   \cmark &   \cmark &   \cmark \\   
225 & \cmark &   \cmark &   \cmark &   \cmark &   \xmark &  \cmark &   \cmark \\   
255 & \cmark &   \cmark &   \cmark &   \cmark &   \xmark &  \cmark &   \cmark \\   
285 & \cmark &   \cmark &   \cmark &   \cmark &   \cmark &   \cmark &   \cmark \\   
300 & \cmark &   \cmark &   \cmark &   \cmark &   \cmark &   \cmark &   \cmark \\
\bottomrule
    \end{tabular}
    }
    \caption{Results of the KS test for the \gls{e2e} latency obtained using the statistical models and real data with the SemanticKITTI dataset. The mark \cmark (\xmark) is used if the test is passed (not passed).}
    \label{tab:ks_lat}
\end{table}

\subsubsection{Model accuracy}

We now formally validate the accuracy of the statistical models for the file size of the point clouds derived in \cref{sec:traffic_analysis} measuring their impact on network metrics.
To do so, we repeat the \gls{ks} test on the \gls{e2e} throughput and latency.
The results are reported in \cref{tab:ks_thr,tab:ks_lat}, respectively.
As expected, the test is passed for almost all compression configurations and distances, so it is another demonstration of the accuracy of the selected models.
The only exception is with D0/S0, which also failed the test in \cref{ssub:statistical-results}, and for $d>150$~m, where propagation is largely in \gls{nlos}.
In this range, the channel is severely unstable and may introduce unpredictable non-linear behaviors, making the statistical models less representative of the real~data.

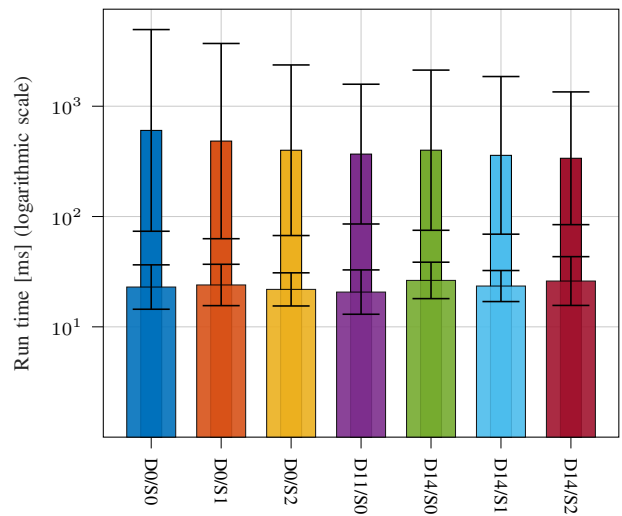
\begin{figure}[t]
    \setlength\fheight{.3\columnwidth}
    \setlength\fwidth{\columnwidth}
    \centering
\begin{tikzpicture}

\definecolor{brown1611946}{RGB}{161,19,46}
\definecolor{chocolate2168224}{RGB}{216,82,24}
\definecolor{darkcyan0113188}{RGB}{0,113,188}
\definecolor{darkgray176}{RGB}{176,176,176}
\definecolor{goldenrod23617631}{RGB}{236,176,31}
\definecolor{lightgray204}{RGB}{204,204,204}
\definecolor{mediumturquoise76189237}{RGB}{76,189,237}
\definecolor{olivedrab11817147}{RGB}{118,171,47}
\definecolor{purple12546141}{RGB}{125,46,141}
\pgfplotsset{every tick label/.append style={font=\scriptsize}}

\begin{axis}[
tick align=outside,
tick pos=left,
x grid style={lightgray204},
xmajorgrids,
xmin=-0.685, xmax=6.685,
xminorgrids,
xtick style={color=black},
xtick={0,1,2,3,4,5,6},
xticklabel style={rotate=270},
xticklabels={D0/S0,D0/S1,D0/S2,D11/S0,D14/S0,D14/S1,D14/S2},
y grid style={lightgray204},
ylabel={Run time [ms] (logarithmic scale)},
ymajorgrids,
ymin=0, ymax=3.87961388713014,
yminorgrids,
ytick style={color=black},
ylabel style={font=\footnotesize\color{white!15!black}},
ytick={1,2,3,4},
yticklabels={$10^1$, $10^2$, $10^3$, $10^4$}
]

\draw[draw=black,fill=darkcyan0113188] (axis cs:-0.15,0) rectangle (axis cs:0.15,2.78069509832341);
\draw[draw=black,fill=chocolate2168224] (axis cs:0.85,0) rectangle (axis cs:1.15,2.68389806406979);
\draw[draw=black,fill=goldenrod23617631] (axis cs:1.85,0) rectangle (axis cs:2.15,2.60092324195305);
\draw[draw=black,fill=purple12546141] (axis cs:2.85,0) rectangle (axis cs:3.15,2.56587729982891);
\draw[draw=black,fill=olivedrab11817147] (axis cs:3.85,0) rectangle (axis cs:4.15,2.6013438329404);
\draw[draw=black,fill=mediumturquoise76189237] (axis cs:4.85,0) rectangle (axis cs:5.15,2.55458568915121);
\draw[draw=black,fill=brown1611946] (axis cs:5.85,0) rectangle (axis cs:6.15,2.52806010002415);

\draw[draw=black,fill=darkcyan0113188,opacity=0.9] (axis cs:-0.35,0) rectangle (axis cs:0.35,1.36072628510067);
\draw[draw=black,fill=chocolate2168224,opacity=0.9] (axis cs:0.65,0) rectangle (axis cs:1.35,1.37964969683174);
\draw[draw=black,fill=goldenrod23617631,opacity=0.9] (axis cs:1.65,0) rectangle (axis cs:2.35,1.33939773126912);
\draw[draw=black,fill=purple12546141,opacity=0.9] (axis cs:2.65,0) rectangle (axis cs:3.35,1.3150800392646);
\draw[draw=black,fill=olivedrab11817147,opacity=0.9] (axis cs:3.65,0) rectangle (axis cs:4.35,1.42087099173855);
\draw[draw=black,fill=mediumturquoise76189237,opacity=0.9] (axis cs:4.65,0) rectangle (axis cs:5.35,1.36980052064824);
\draw[draw=black,fill=brown1611946,opacity=0.9] (axis cs:5.65,0) rectangle (axis cs:6.35,1.41511855465979);

\path [draw=black, semithick]
(axis cs:0,1.15948827371471)
--(axis cs:0,1.56196429648663);

\addplot [semithick, black, mark=-, mark size=7, mark options={solid}, only marks]
table {%
0 1.15948827371471
};
\addplot [semithick, black, mark=-, mark size=7, mark options={solid}, only marks]
table {%
0 1.56196429648663
};
\path [draw=black, semithick]
(axis cs:1,1.19220931469832)
--(axis cs:1,1.56709007896516);

\addplot [semithick, black, mark=-, mark size=7, mark options={solid}, only marks]
table {%
1 1.19220931469832
};
\addplot [semithick, black, mark=-, mark size=7, mark options={solid}, only marks]
table {%
1 1.56709007896516
};
\path [draw=black, semithick]
(axis cs:2,1.189171625886)
--(axis cs:2,1.48962383665223);

\addplot [semithick, black, mark=-, mark size=7, mark options={solid}, only marks]
table {%
2 1.189171625886
};
\addplot [semithick, black, mark=-, mark size=7, mark options={solid}, only marks]
table {%
2 1.48962383665223
};
\path [draw=black, semithick]
(axis cs:3,1.11409080383157)
--(axis cs:3,1.51606927469764);

\addplot [semithick, black, mark=-, mark size=7, mark options={solid}, only marks]
table {%
3 1.11409080383157
};
\addplot [semithick, black, mark=-, mark size=7, mark options={solid}, only marks]
table {%
3 1.51606927469764
};
\path [draw=black, semithick]
(axis cs:4,1.2557307918588)
--(axis cs:4,1.58601119161831);

\addplot [semithick, black, mark=-, mark size=7, mark options={solid}, only marks]
table {%
4 1.2557307918588
};
\addplot [semithick, black, mark=-, mark size=7, mark options={solid}, only marks]
table {%
4 1.58601119161831
};
\path [draw=black, semithick]
(axis cs:5,1.22910581153523)
--(axis cs:5,1.51049522976125);

\addplot [semithick, black, mark=-, mark size=7, mark options={solid}, only marks]
table {%
5 1.22910581153523
};
\addplot [semithick, black, mark=-, mark size=7, mark options={solid}, only marks]
table {%
5 1.51049522976125
};
\path [draw=black, semithick]
(axis cs:6,1.19403342872699)
--(axis cs:6,1.63620368059258);

\addplot [semithick, black, mark=-, mark size=7, mark options={solid}, only marks]
table {%
6 1.19403342872699
};
\addplot [semithick, black, mark=-, mark size=7, mark options={solid}, only marks]
table {%
6 1.63620368059258
};
\path [draw=black, semithick]
(axis cs:0,1.86651982795145)
--(axis cs:0,3.69487036869537);

\addplot [semithick, black, mark=-, mark size=7, mark options={solid}, only marks]
table {%
0 1.86651982795145
};
\addplot [semithick, black, mark=-, mark size=7, mark options={solid}, only marks]
table {%
0 3.69487036869537
};
\path [draw=black, semithick]
(axis cs:1,1.79898373387964)
--(axis cs:1,3.56881239425994);

\addplot [semithick, black, mark=-, mark size=7, mark options={solid}, only marks]
table {%
1 1.79898373387964
};
\addplot [semithick, black, mark=-, mark size=7, mark options={solid}, only marks]
table {%
1 3.56881239425994
};
\path [draw=black, semithick]
(axis cs:2,1.82762582538944)
--(axis cs:2,3.37422065851666);

\addplot [semithick, black, mark=-, mark size=7, mark options={solid}, only marks]
table {%
2 1.82762582538944
};
\addplot [semithick, black, mark=-, mark size=7, mark options={solid}, only marks]
table {%
2 3.37422065851666
};
\path [draw=black, semithick]
(axis cs:3,1.93234436442959)
--(axis cs:3,3.19941023522824);

\addplot [semithick, black, mark=-, mark size=7, mark options={solid}, only marks]
table {%
3 1.93234436442959
};
\addplot [semithick, black, mark=-, mark size=7, mark options={solid}, only marks]
table {%
3 3.19941023522824
};
\path [draw=black, semithick]
(axis cs:4,1.87526202119569)
--(axis cs:4,3.32742564468511);

\addplot [semithick, black, mark=-, mark size=7, mark options={solid}, only marks]
table {%
4 1.87526202119569
};
\addplot [semithick, black, mark=-, mark size=7, mark options={solid}, only marks]
table {%
4 3.32742564468511
};
\path [draw=black, semithick]
(axis cs:5,1.840037283245)
--(axis cs:5,3.26913409505743);

\addplot [semithick, black, mark=-, mark size=7, mark options={solid}, only marks]
table {%
5 1.840037283245
};
\addplot [semithick, black, mark=-, mark size=7, mark options={solid}, only marks]
table {%
5 3.26913409505743
};
\path [draw=black, semithick]
(axis cs:6,1.92637610070241)
--(axis cs:6,3.12974409934588);

\addplot [semithick, black, mark=-, mark size=7, mark options={solid}, only marks]
table {%
6 1.92637610070241
};
\addplot [semithick, black, mark=-, mark size=7, mark options={solid}, only marks]
table {%
6 3.12974409934588
};
\end{axis}

\end{tikzpicture}
    \caption{Average run time for different \gls{hsc} compression configurations measured in ns-3 on real data from the SemanticKITTI dataset (narrow bars) and using the statistical traffic models (wide bars).}
    \label{fig:runtime}
\end{figure}

\subsubsection{Run time}
Finally, in \cref{fig:runtime} we report the run time of a single simulations run in ns-3 using either the \textit{BurstyApplication}, i.e., real data from the SemanticKITTI dataset (narrow bars), or the \textit{StatisticalTraffic} application, which relies on statistical models (wide bars).
As expected, the run time improves when using statistical models, with an average speed-up factor of $18\times$.
The maximum gain is observed with raw data (D0/S0), where the simulation time decreases from over 10 minutes to approximately 22 seconds ($26 \times$ speed-up).

In conclusion, the \textit{StatisticalTraffic} application proves to be a valid alternative to the \textit{BurstyApplication}, delivering substantial improvements in simulation run time while maintaining high accuracy. The only restriction is for $d>150$~m, where our statistical models may not fully capture the characteristics of real data. However, this limitation is negligible since direct V2X communication at these distances is generally impractical, if not entirely infeasible, under realistic latency constraints.

\section{Conclusion}
\label{sec:conclusions}
In this paper we proposed a comprehensive statistical characterization for the size of LiDAR point clouds based on the SemanticKITTI dataset, to be used for network simulations in an automotive driving scenario.
We obtained seven distinct distributions to model raw data and six representative compression configurations using the state-of-the-art HSC algorithm. The statistical models have been rigorously validated through a Kolmogorov-Smirnov test with Bootstrap Resampling, with the exception of the model representing raw (uncompressed) data that did not pass the test.
Furthermore, we implemented these random distributions in ns-3, and ran a simulation campaign to evaluate their accuracy in terms of some \gls{e2e} network metrics.
Our results showed a perfect match between our statistical traffic models and real data from SemanticKITTI. 
We also observed that compression can significantly improve the network throughput, and highlighted the critical trade-off between compression ratio and latency.
Additionally, we found that using statistical models substantially improves the simulation run time compared to using real data, while maintaining high accuracy. However, the reliability decreases for distances beyond 150 m, where channel conditions become statistically unstable.

As part of future work, we will leverage the flexibility of HSC to further refine our models, incorporating additional factors such as the impact of compression on data quality and energy consumption, besides performance metrics such as throughput and latency.


\bibliographystyle{IEEEtran}
\bibliography{bibl.bib}
\vspace{12pt}
\color{red}

\end{document}